# Phase Transitions in Warm, Asymmetric Nuclear Matter

Horst Müller and Brian D. Serot
*Physics Department and Nuclear Theory Center*
*Indiana University, Bloomington, Indiana 47405*
(May 5, 1995)

## Abstract

A relativistic mean-field model of nuclear matter with arbitrary proton fraction is studied at finite temperature. An analysis is performed of the liquid–gas phase transition in a system with two conserved charges (baryon number and isospin) using the stability conditions on the free energy, the conservation laws, and Gibbs' criteria for phase equilibrium. For a binary system with two phases, the coexistence surface (binodal) is two-dimensional. The Maxwell construction through the phase-separation region is discussed, and it is shown that the stable configuration can be determined uniquely at every density. Moreover, because of the greater dimensionality of the binodal surface, the liquid–gas phase transition is continuous (second order by Ehrenfest's definition), rather than discontinuous (first order), as in familiar one-component systems. Using a mean-field equation of state calibrated to the properties of nuclear matter and finite nuclei, various phase-separation scenarios are considered. The model is then applied to the liquid–gas phase transition that may occur in the warm, dilute matter produced in energetic heavy-ion collisions. In asymmetric matter, instabilities that produce a liquid–gas phase separation arise from fluctuations in the proton concentration (chemical instability), rather than from fluctuations in the baryon density (mechanical instability).

## I. INTRODUCTION

The determination of the properties of nuclear matter as functions of density, temperature, and the ratio of protons to neutrons is a fundamental problem in nuclear physics. To achieve this goal, one must study not only the ground and excited states of normal nuclei, but also highly excited nuclei created in nucleus–nucleus collisions and nuclei far from stability, which may be created in radioactive beams. In this work we consider the properties of equilibrium nuclear matter at finite temperature and arbitrary proton fraction, and we describe some new qualitative features that may be relevant for the energetic collisions of heavy ions.

There have been numerous theoretical studies of the dynamics of medium-energy heavy-ion collisions [1–6]. Some of these are based on equilibrium thermodynamics and focus on the nuclear matter phase diagram [7–11]. The basic feature is the liquid–gas phase transition at low densities and moderate temperatures, and how the nuclear system evolves through various phase-separation boundaries (*binodals*) and instability boundaries (*spinodals*) [7,12,13]. Other analyses concentrate on the nonequilibrium evolution of the system by describing the early stages of the collision through cascade simulations [14,15], for example, or the late stages of the collision using models of fragmentation or nucleation [16–18]. Although the equilibrium analysis oversimplifies this problem and is applicable only to certain aspects of the evolution, it is useful for providing an orientation to the complex dynamics by giving concrete descriptions of the phase-separation process, which may be difficult to characterize in more microscopic formulations. Here we will follow the thermodynamic approach and focus on the qualitatively new features that arise when a system with two conserved charges (*i.e.*, baryon number and electric charge) undergoes a liquid–gas phase transition. These features have been discussed earlier by Barranco and Buchler [19], who use a simple phenomenological equation of state with an interaction energy that is temperature independent, and by Glendenning [20], who focuses on zero-temperature matter and neutron stars. More recently [21], Pethick, Ravenhall, and Lorenz used a similar analysis to study the composition of neutron star crusts.

The main ingredient in the analysis is the nuclear equation of state, and ours is based on a relativistic mean-field model involving the interaction of Dirac nucleons with scalar and vector mesons [22,23]. There are several reasons for choosing this type of model. First, recent developments in the application of these models to the structure of nuclei show that they can provide an excellent description of bulk nuclear properties throughout the periodic table, provided that nonlinear scalar and vector self-interactions are included [23–34]. In fact, relativistic mean-field models describe these properties as well as or better than [35] any other microscopic model presently available. Thus we have a way to calibrate our equation of state at zero temperature and normal nuclear densities, and then extrapolate into the warm, dilute regime appropriate for the phase transition. Second, the mean-field approximation is known to be thermodynamically consistent; that is, it satisfies the relevant thermodynamic identities and the virial theorem [36]. Moreover, although the mean-field approximation oversimplifies the nuclear dynamics, it allows for easy computations and is commensurate with our simplified description of the collision dynamics. Finally, the ability to determine the model parameters analytically from a specified set of zero-temperature nuclear matter properties [30,34] allows us to easily study variations of the model and to assess the sensitivity of our results to the nuclear compressibility and symmetry energy, both of which are not particularly well known.

The models we study involve nonlinear scalar and vector self-interactions through fourth order in these fields, as first proposed by Boguta and Bodmer [24], and later generalized by Bodmer and Price [28,30]. The desired nuclear symmetry energy is achieved using the simplest possible coupling of the $\rho$ meson to the nucleon, and there are no $\rho$ self-interactions. Although it is known that this way of generating the symmetry energy is a simplification of more complex mechanisms (for example, one-pion-exchange in a Hartree–Fock calculation produces significant contributions [37]), and that the dependence of the energy on the isovector density ($\rho_3 \equiv \rho_{\mathrm{p}} - \rho_{\mathrm{n}}$) may be more complicated when one is far from symmetric matter, the qualitative features studied here arise because the symmetry energy is repulsive for *all* values of $\rho_3$, and this repulsion grows monotonically as $|\rho_3|$ increases. Thus the present simple description is adequate for our purposes, and we expect that other models of the symmetry energy with these features will produce similar results. Moreover, we can adjust the symmetry energy to any reasonable value, and it is straightforward to generalize our model, which we leave as a topic for future investigations.

Similar relativistic mean-field calculations of warm, *symmetric* nuclear matter were performed earlier by one of us [36], but there the original Walecka model [38] was used, which allows only a rough calibration to observed bulk nuclear properties. Here we want to explore arbitrary proton fractions and to present results for the best relativistic equations of state available. Moreover, this earlier work concentrated on the Lorentz covariance of the relativistic mean-field theory; the same analysis could be applied to the class of models studied here, but this is unnecessary, and we carry out all calculations in the rest frame of the warm matter. Indeed, even the use of a relativistic mean-field theory is not essential for our discussion; any equation of state that is accurately calibrated to nuclei should produce similar results.

We emphasize that although we present quantitative results based on an accurate nuclear equation of state, our focus is on the new qualitative features that arise in liquid–gas phase transitions in systems with more than one conserved charge. We believe that our simplified hydrodynamic, thermodynamic, and mean-field description of nucleus–nucleus collisions is the most transparent way to do this, and we expect that signatures of these new features will survive in more sophisticated calculations. In particular, our primary result is that the liquid–gas phase transition in asymmetric nuclear matter is of *second order*, in contrast to the familiar first-order (van der Waals) transition that occurs in one-component systems (and in symmetric nuclear matter). If such a phase transition actually occurs in medium-energy heavy-ion collisions, then generically it should be smoother than a first-order phase transition. To be more precise, even if the assumption of thermodynamic equilibrium is totally justified, and even if the finite-size effects on the phase separation are totally negligible, we show that the generic liquid–gas phase transition *must be continuous*; for example, it occurs over a temperature range of several MeV for matter that is roughly 40% protons. Thus observable signals of this transition should be more continuous than previously expected from conventional analyses, which find a discontinuous (first-order) phase transition in the thermodynamic limit [16].

Furthermore, we find that in asymmetric matter, the relevant spinodal that signals instability to infinitesimal fluctuations arises from variations in the proton concentration at fixed pressure. This is in contrast to the usual scenario (and the one relevant for symmetric matter), in which spinodal decomposition occurs due to fluctuations in the baryon density at points of mechanical instability. Thus microscopic models that attempt to describe this decomposition using nucleation or fragmentation must be generalized to allow for different proton concentrations in the liquid and gas phases.

The outline of this paper is as follows: In Sec. II, we present a general discussion of the thermodynamics of phase transitions in multicomponent systems. We emphasize that the stability conditions, conservation laws, and Gibbs' criteria produce a set of equations that completely specifies the thermodynamic variables. (That is, there are equal numbers of equations and unknowns.) While some of this material has been presented elsewhere, we believe this is the first derivation that proceeds directly from a study of the free energy of the system. In Sec. III, we describe the relativistic mean-field model, present the relations that determine the equation of state, and discuss how the model parameters are specified. Some numerical procedures used to obtain our results are also discussed. Section IV deals with a study of the binodal surface; various phase diagrams and Maxwell constructions are illustrated, and the evidence for the second-order liquid–gas transition is presented. In Sec. V, we apply our simple model to heavy-ion collisions and describe the phase coexistence region and the various spinodals. We also consider how the continuous phase transition may affect the evolution of the system. Section VI contains a short summary.

As a guide for the reader, we note that the material in Secs. II through IV is presented before the applications in Sec. V for both logical and pedagogical reasons. Nevertheless, because of the formal nature of the initial discussion, it may be useful on a first reading to consider Secs. II through IV briefly and then to concentrate on the results of the formalism that are displayed in Sec. V. After gaining some insight into the behavior of a specific two-component system, one can then return to the formal material and consider it in more depth.

## II. PHASE TRANSITIONS IN MULTICOMPONENT SYSTEMS

We consider a system characterized by a hamiltonian $\hat{H}$ and a set of $n$ mutually commuting charges $\hat{Q}_i$. Here a conserved charge does not necessarily imply an independent particle species; it includes any conserved quantity resulting from an underlying symmetry of the system, for example, electric charge, baryon number, total angular momentum, *etc.* Moreover, particle numbers may change during a process (by decay, for example), and it is actually the conserved *charges* that are relevant for the thermodynamic analysis.

The equilibrium state of the system enclosed in a volume $V$ is completely described by the thermodynamic potential [39]

$$\Omega(T,V,\mu_i) = -\frac{1}{\beta}\ln \mathrm{Tr}\exp\Big(-\beta(\hat{H} - \sum_i \mu_i \hat{Q}_i)\Big) \ , \tag{1}$$

where $\beta$ is the inverse temperature. Thus the average charges and the pressure can be obtained from

$$Q_j(T,V,\mu_i) = -\left(\frac{\partial\Omega}{\partial\mu_j}\right)_{T,V,\{\mu_i,i\neq j\}} \ , \tag{2}$$

$$p(T,V,\mu_i) = -\left(\frac{\partial\Omega}{\partial V}\right)_{T,\{\mu_i\}} \ . \tag{3}$$

In some regions of $T$ and $\mu_i$, several choices of the $\mu_i$ may lead to the same $p$, and this allows for the possibility of phase transitions. To be more specific, let us first perform a Legendre transformation and consider the Helmholtz free energy $F$. We assume a system with specified charges $Q_i$ in a large volume $V$, so that surface effects can be neglected. We then have

$$\Omega(T,V,\mu_i) = -Vp(T,\mu_i) \ , \tag{4}$$

$$Q_i \equiv V\rho_i(T,\mu_i) \ , \tag{5}$$

and correspondingly,

$$F(T,V,Q_i) \equiv V\mathcal{F}(T,\rho_i) = \Omega(T,V,\mu_i) + V\sum_i \mu_i\rho_i(T,\mu_i) \ , \tag{6}$$

with

$$\mu_j = \left(\frac{\partial\mathcal{F}(T,\rho_i)}{\partial\rho_j}\right)_{T,\{\rho_i,i\neq j\}} \ . \tag{7}$$

The system will be stable against separation into two phases if the free energy of a single phase is lower than the free energy in all two-phase configurations. This requirement can be formulated as

$$\mathcal{F}(T,\rho_i) < (1-\lambda)\mathcal{F}(T,\rho_i') + \lambda\mathcal{F}(T,\rho_i'') \ , \tag{8}$$

with

$$\rho_i = (1-\lambda)\rho_i' + \lambda\rho_i'' \ , \qquad 0 < \lambda < 1 \ , \tag{9}$$

where the two phases are denoted by a prime and a double-prime. In formal terms, stability implies that the free energy density is a *convex* function of the densities $\rho_i$ [40]. Convexity implies that stability against separation into two phases also guarantees stability against separation into an arbitrary number of phases (which we denote by superscripts $\alpha$, $\beta$, ..., $\omega$). That is, by Jensen's inequality on convex functions, Eqs. (8) and (9) are equivalent to

$$\mathcal{F}(T,\rho_i) < \sum_{\{\alpha\}} \lambda_\alpha \mathcal{F}(T,\rho_i^\alpha) \ , \tag{10}$$

with

$$\rho_i = \sum_{\{\alpha\}} \lambda_\alpha \rho_i^\alpha \quad \text{and} \quad \sum_{\{\alpha\}} \lambda_\alpha = 1 \ , \tag{11}$$

where the sums are over all phases $\{\alpha\}$. The parameters $\lambda^\alpha = V^\alpha/V$ specify the volume fraction occupied by each phase. The second equation (9) or (11) ensures that the overall charges are conserved:

$$V\rho_i = \sum_{\{\alpha\}} V^\alpha \rho_i^\alpha \ , \tag{12}$$

$$\text{with} \qquad V = \sum_{\{\alpha\}} V^\alpha \ . \tag{13}$$

Equations (8) and (10) are *global* criteria for the stability of the one-phase system. If these inequalities are satisfied, then it is necessarily true that the symmetric matrix

$$\mathcal{F}_{ij} \equiv \left(\frac{\partial^2 \mathcal{F}}{\partial\rho_i \partial\rho_j}\right)_T$$

is positive [40]. In contrast, whenever Eq. (8) is violated, a system with more than one phase is energetically favorable. The phase coexistence is governed by the Gibbs' conditions

$$\mu_i^\alpha(T,\rho_i^\alpha) = \mu_i^\beta(T,\rho_i^\beta) = \cdots = \mu_i^\omega(T,\rho_i^\omega) \ , \quad i = 1,\ldots,n \ , \tag{14}$$

$$p^\alpha(T,\rho_i^\alpha) = p^\beta(T,\rho_i^\beta) = \cdots = p^\omega(T,\rho_i^\omega) \ , \tag{15}$$

where the temperature is the same in all phases. The *local* positivity conditions on $\mathcal{F}_{ij}$ give rise to a set of $n$ inequalities that divide the parameter space $\{T,\rho_i\}$ into stable and unstable regions. It is important to realize, however, that under conditions of phase separation, Eq. (8) may still be valid *locally* (that is, for $\rho_i', \rho_i'' \approx \rho_i$), but it may nevertheless be possible to find significantly different densities $\rho_i'$ and $\rho_i''$ that violate this condition. This leads to the existence of metastable states.

These ideas are illustrated for a simple one-component system in Fig. 1. Here the free energy density at fixed temperature is shown as a function of the density. At point $D$, we evidently have

$$\frac{\partial^2 \mathcal{F}}{\partial\rho^2} > 0 \ ,$$

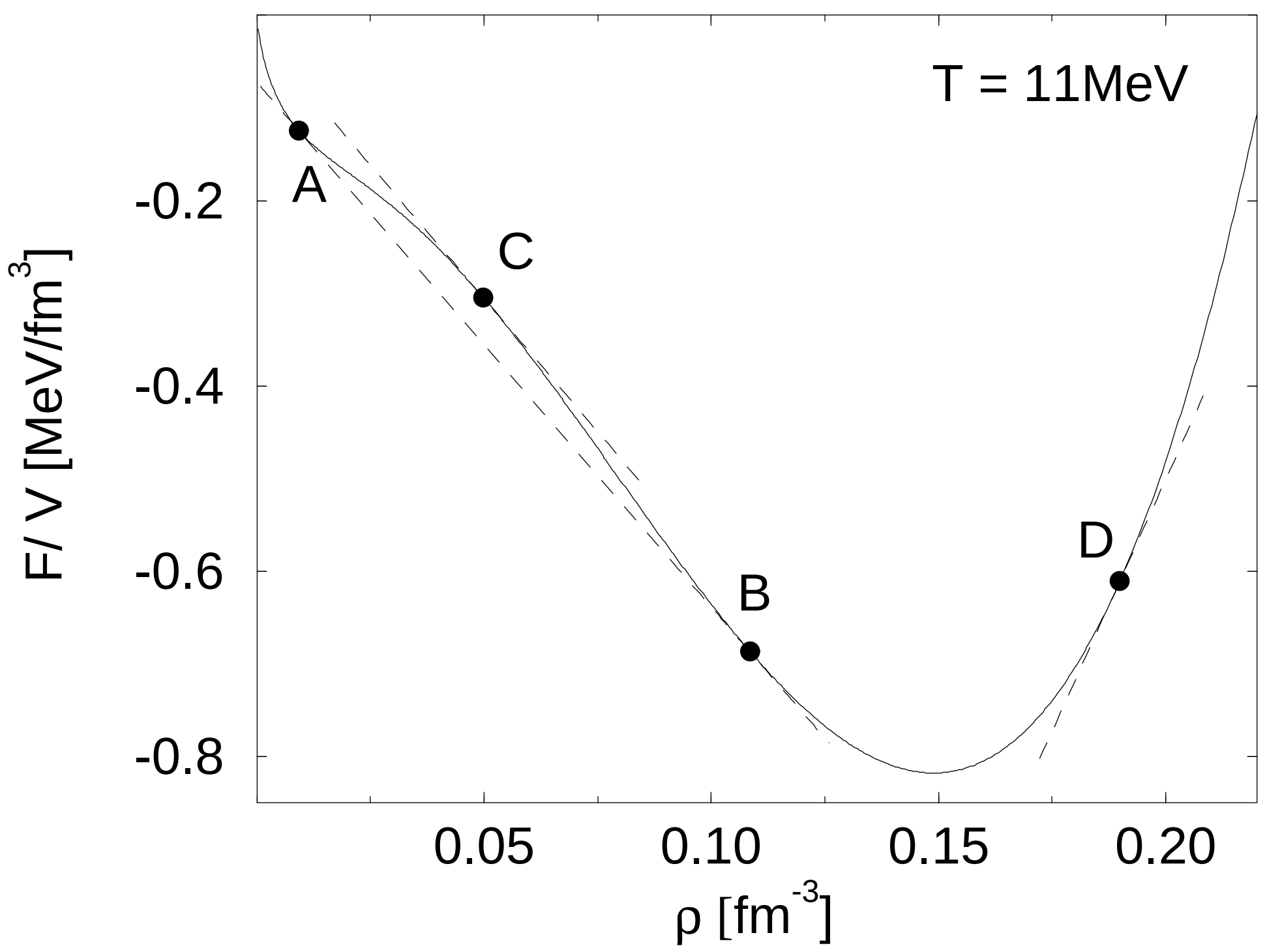

FIG. 1. Free-energy density as a function of density for a one-component system at fixed temperature. For clarity, we show $\mathcal{F} - e_0\rho \equiv \mathcal{F} + (15.75\,\mathrm{MeV})\rho$, since the variations in $\mathcal{F}$ are quite small. The additional term linear in the density has no effect on the stability conditions or on Gibbs' criteria.

and the system is stable. At points $A$ and $B$, $\mathcal{F}$ is still locally convex, but these points share a common tangent (chemical potential). Thus, although the stability criteria remain valid locally around $A$ and $B$, the existence of the common tangent implies that these two phases with different densities can coexist. The values of $T$ and $\rho$ at the points $A$ and $B$ lie on the binodal surface in the $\{T, \rho\}$ plane, which is a one-dimensional surface in this simple example. At point $C$, the free energy is concave, indicating that this configuration is unstable against phase separation. For points between $A$ and $B$ where the second derivative of $\mathcal{F}$ remains positive, the system is metastable and can exist temporarily in a single phase, allowing for supercooling or superheating. However, there is an inflection point between $A$ and $C$, and also one between $B$ and $C$, where the second derivative of $\mathcal{F}$ vanishes; these points lie on the spinodal surface, which delimits the onset of instability.

According to Gibbs' phase rule [41], at most

$$K_{\max} = n + 2$$

phases can coexist in equilibrium in a system with $n$ conserved charges. Each single phase is characterized by the set $(T, \rho_i^\alpha, V^\alpha)$ of $n+2$ variables, so that if there are $K$ phases, the total number of variables is $K(n+1)+1$. (All phases have the same temperature.) We assume that the $n$ total charges $Q_i$ and the total volume $V$ are also specified, which brings the total number of variables to $K(n+1)+n+2$. These variables are restricted by the Gibbs' conditions (14) and (15), which provide $(K-1)(n+1)$ constraints, and by the conservation laws (11) on the charges ($n$ constraints) and (13) on the total volume (1 constraint). We therefore end up with a set of $n+2$ independent variables, which we can take to be $(T, Q_i, V)$ or $(T, \rho_i, V)$. This result implies that the number of degrees of freedom in a system with strictly conserved charges is independent of the number of phases [42,43].

It is important to emphasize, however, an important difference between a multicomponent system and one containing a single conserved charge. Although the ratios of the total charges $Q_i/Q_j$ remain fixed once the system has been prepared, *the ratios can be different in different phases.* Moreover, since the independent variables $T$, $Q_i$, and $V$ determine the energetically stable state of the system, it is impossible to impose additional constraints. For example, as shown in Ref. [20], one cannot demand that the pressure remain constant during a phase transition at fixed temperature, for this would violate either the equilibrium conditions or the conservation laws (12). Consequently, the common (vapor) pressures and chemical potentials vary when the proportions $\lambda^\alpha$ of the phases change. These results will be illustrated explicitly for the case of asymmetric nuclear matter in Secs. IV and V.

A necessary condition for the coexistence of more than two phases is that any pair of these phases must be in equilibrium. In addition, as mentioned earlier, a system that is stable against separation into two phases is also stable against separation into multiple phases. Let us therefore focus on two such phases for a moment, which we will denote with a prime and a double-prime. The two sets of densities $\{\rho_i', \rho_i''\}$ that satisfy Eqs. (14) and (15) form a surface in the parameter space $\{T, \rho_i\}$; this is the phase separation boundary, or binodal. For $n$ conserved charges and two coexisting phases, Gibbs' phase rule implies that the binodal is an $n$-dimensional surface. We can also show that this surface encloses all points satisfying

$$\rho_i = (1-\lambda)\rho_i' + \lambda\rho_i'' \ , \qquad 0 < \lambda < 1 \ , \tag{16}$$

that lead to a single (unstable) configuration with a higher value for the free energy. Although this statement is reasonable from a physical point of view, it is not immediately obvious.

To prove it, we define a tangent plane $\mathbf{T}_{\{\bar{\rho}_i\}}(\rho_i)$ that is a function of the $\rho_i$ and that can be attached to any point in the parameter space $\bar{\rho}_i$. Since the slopes at any such point are given by the chemical potentials $\mu_i(\bar{\rho}_i)$ [see Eq. (7) and Fig. 1], these tangent planes can be expressed as

$$\mathbf{T}_{\{\bar{\rho}_i\}}(\rho_i) = \sum_{i=0}^{n} (\rho_i - \bar{\rho}_i)\mu_i(\bar{\rho}_i) + \mathcal{F}(T, \bar{\rho}_i) = \sum_{i=0}^{n} \rho_i\mu_i(\bar{\rho}_i) - p(T, \bar{\rho}_i) \ . \tag{17}$$

From the first equality, it follows immediately that when $\mathcal{F}(T, \rho_i)$ is a convex function of the $\rho_i$, the tangent plane will lie below $\mathcal{F}$ (the stable region), and when the convexity of $\mathcal{F}$ is lost (that is, when $\mathcal{F}$ acquires a "saddle point"), the tangent plane will lie above $\mathcal{F}$ (the unstable region) [40]. Moreover, from the second equality, the existence of a common tangent plane connecting two distant configurations implies equal pressures and chemical potentials in these configurations, which are simply Gibbs' criteria for phase equilibrium. These common tangent planes therefore define a binodal surface and allow for phase separation; it also follows that every point in the parameter space that leads to an unstable single-phase system must lie inside the region enclosed by the binodal surface. This is evident in Fig. 1 for the unstable systems in the neighborhood of point $C$, which lie inside the binodal surface that contains $A$ and $B$.

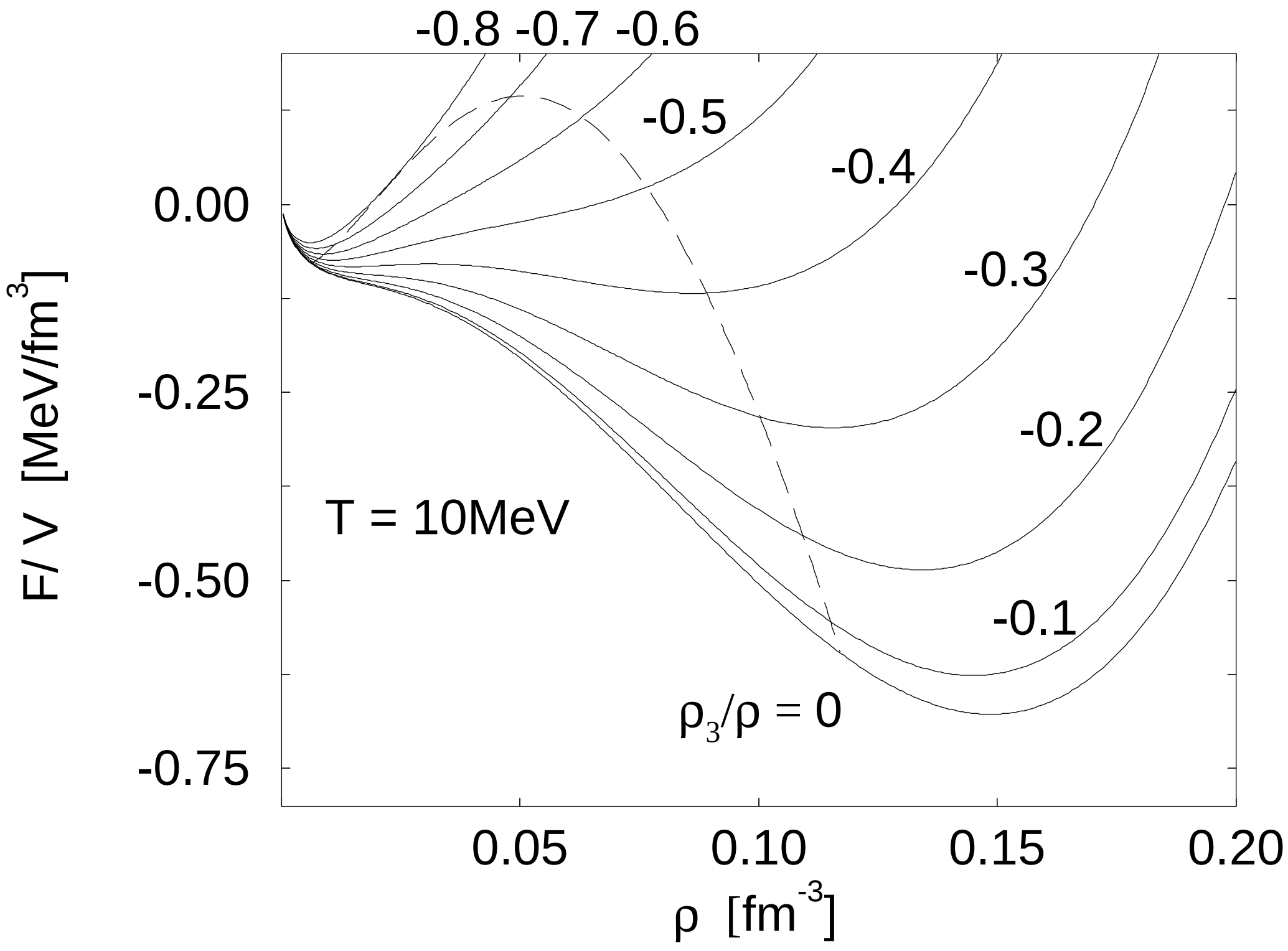

FIG. 2. Free-energy density as a function of density and asymmetry $\rho_3/\rho$ for a system with two conserved charges at fixed temperature. We again show $\mathcal{F} - e_0\rho$ for clarity. The curve for symmetric matter has $\rho_3/\rho = 0$.

In Fig. 2, we show the corresponding situation for a system with two conserved charges at fixed temperature. Here $\rho$ denotes the density of the sum of the charges, and $\rho_3$ denotes the density of the difference; thus, $\rho_3/\rho$ is a measure of the asymmetry. The dashed curve indicates the intersection of the two-dimensional binodal surface obtained from the set of common tangent planes with the plane defined by $T = 10\,\mathrm{MeV}$. (The endpoints of the dashed curve in Fig. 2 correspond to points $A$ and $B$ in Fig. 1.) Observe that all configurations where the free-energy density has a saddle point are contained within the binodal. Note also that in this example, the free-energy density is always convex with respect to variations in $\rho_3$ at fixed $\rho$ and $T$.

One feature of the binodal surface is that it may contain *critical points.* At the critical points, if they exist, the two phases can no longer be distinguished by their densities. Therefore the critical points form a line that divides the binodal surface into different regions describing either a high density (liquid) or a low density (gas) phase. Finally, we note that more than two phases can coexist if and only if each pair of phases form a binodal, and if all these binodals have a common region of intersection [41].

The binodal surface determines the stability boundaries of the system, but it remains to show how the system behaves inside, *i.e.*, how to interpolate within the metastable and unstable regions using a Maxwell construction. To explain this in more detail, we consider an isothermal compression in a situation where the system can separate into two phases.

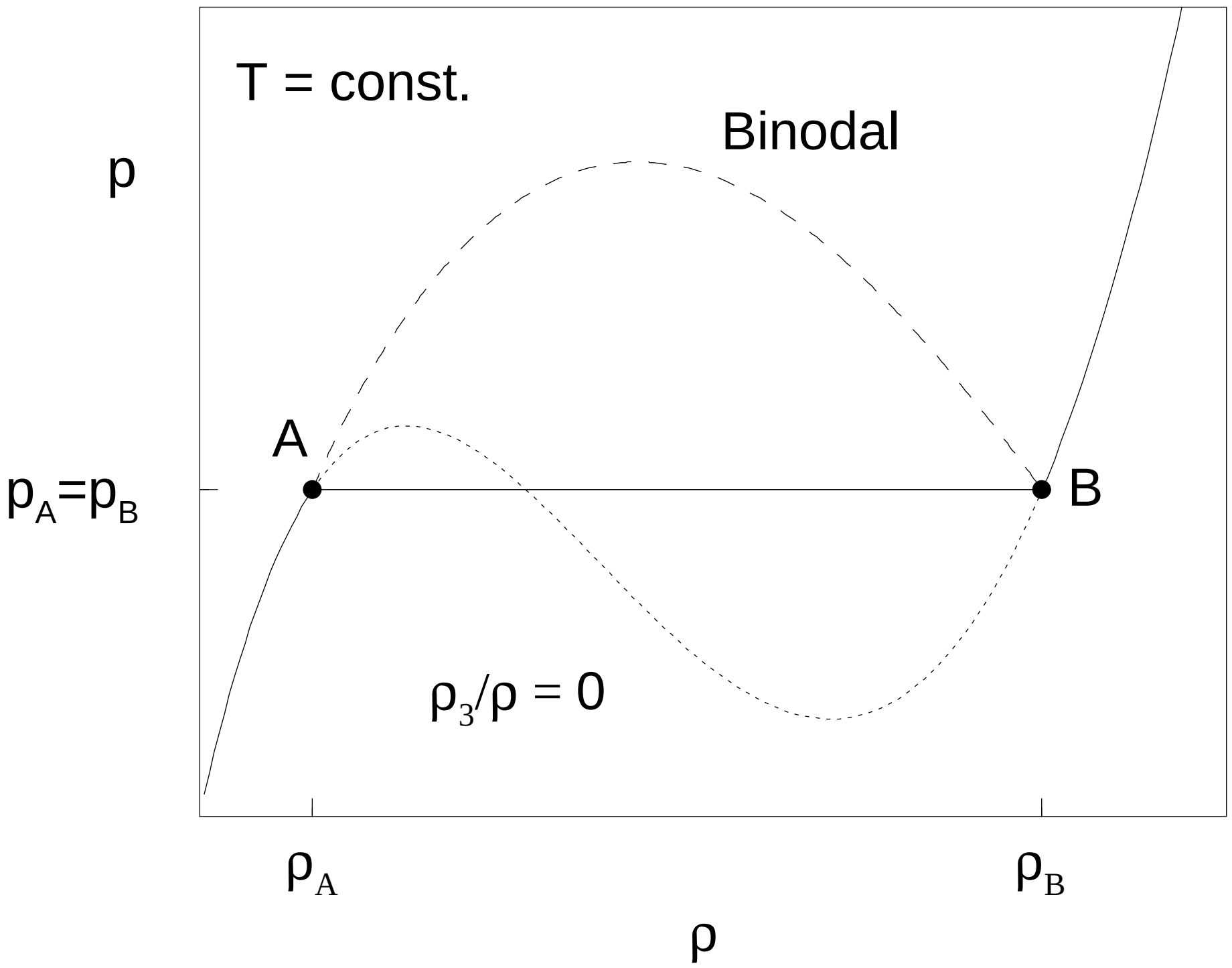

FIG. 3. The Maxwell construction in a one-component system. The binodal line is obtained from similar curves at other temperatures.

We begin with the familiar case of a one-component system, as shown in Fig. 3. Suppose that during the compression, the system encounters the binodal at some point $A$ in $\{T, \rho\}$ space. At this point, the whole volume is occupied by a phase with density $\rho^A$, and a second phase with density $\rho^B$ is about to emerge in an infinitesimally small volume. The two phases at $A$ and $B$ are connected by the Gibbs' conditions, so that they have equal temperatures, pressures, and chemical potentials. In a one-component system, $B$ is the point at which the system leaves the two-phase region; on the $(P, \rho)$ diagram of Fig. 3, $A$ and $B$ are connected by a horizontal line, the well-known Maxwell construction. In a multicomponent system, however, as depicted in Fig. 4, the ratios of the charges in the emerging phase at $B$ are generally different from those in the original preparation at $A$, thus violating the conservation laws. The system must therefore evolve instead through configurations that maintain the ratios of the total charges (the curve $AD$), and it leaves the instability region at the point $D$, which lies together with $A$ on the line of constant ratios $Q_i/Q_j$. At this point, the original phase is present in infinitesimal quantities with densities $\{\rho_i^C\}$, while the newly created phase has evolved to point $D$. The configuration at $D$ is consistent with the conservation laws, and in general, the pressure and chemical potentials in the coexisting phases have changed throughout the transition.

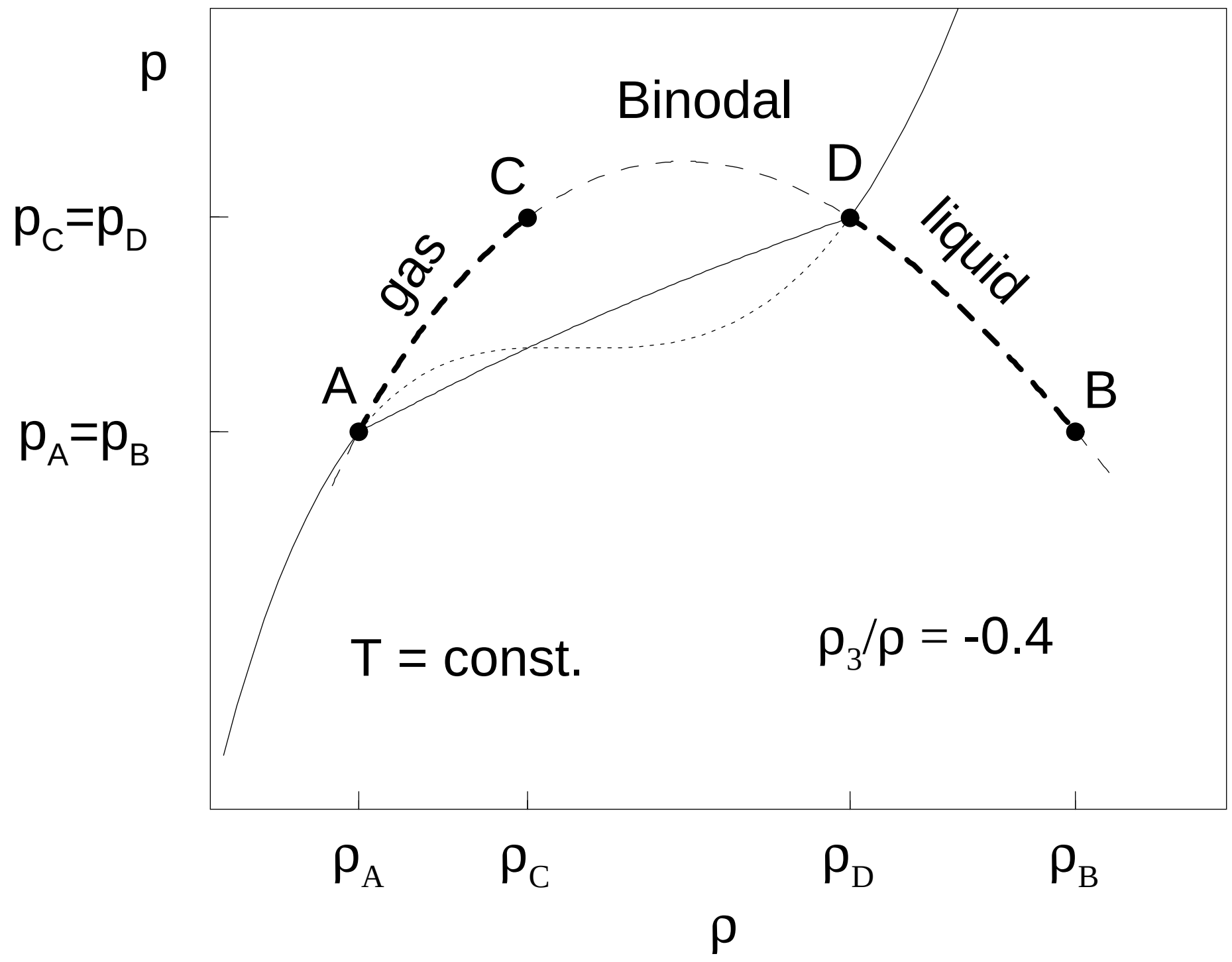

FIG. 4. The Maxwell construction in a two-component system. The asymmetry is held constant at $\rho_3/\rho = -0.4$ throughout the phase separation. Note that only the solid and dotted curves have $\rho_3/\rho = -0.4$.

To determine the nature of the system between these extreme values, we must solve

$$\rho_i = (1-\lambda)\rho_i' + \lambda\rho_i'' \tag{18}$$

for given values of $\rho_i$, with $\rho_i'$ and $\rho_i''$ lying on the binodal surface. It is important to realize that Eqs. (18) are indeed a set of $n$ equations in $n$ unknowns, since the $\rho_i$ are specified, and among the $2n+1$ variables $\rho_i'$, $\rho_i''$, and $\lambda$, $n+1$ can be eliminated by virtue of the Gibbs' conditions (14) and (15). Moreover, these equations yield solutions with qualitatively different characteristics. If the solutions yield all values of $\lambda$ in the interval $[0,1]$, so that

$$\begin{aligned} \rho_i' &= \rho_i^A \qquad \text{for} \qquad \lambda = 0 \ , \\ \rho_i'' &= \rho_i^D \qquad \text{for} \qquad \lambda = 1 \ , \end{aligned}$$

then the system has undergone a phase transition. However, anticipating the subsequent discussion, there are also solutions with $0 \le \lambda \le \lambda_{\max} < 1$. In this case the system becomes unstable to phase separation, but undergoes a *retrograde condensation*: after occupying a maximal volume fraction $\lambda_{\max}$, the new phase begins to disappear, and the system leaves the instability region in the original phase. In either situation, Eq. (18) provides the desired Maxwell construction that determines the free energy in the transition region according to

$$\mathcal{F}(T,\rho_i) = (1-\lambda)\mathcal{F}(T,\rho_i') + \lambda\mathcal{F}(T,\rho_i'') \ .$$

Densities related to other extensive quantities can be computed accordingly.

We close this section by specializing the general formalism to asymmetric nuclear matter, a system of interacting neutrons, protons, and mesons. Such a system is characterized by two conserved charges: the total number of baryons

$$B = N_p + N_n \equiv V\rho \tag{19}$$

and the total charge, or equivalently, the third component of isospin

$$I_3 = \frac{N_p - N_n}{2} \equiv \frac{1}{2} V\rho_3 \ . \tag{20}$$

Thus we have

$$\mathcal{F}(T, \rho, \rho_3) = -p(T, \mu, \mu_3) + \mu\rho + \mu_3\rho_3 \ . \tag{21}$$

The stability condition (8) implies the following set of inequalities on the convex free energy density:

$$\frac{\partial^2 \mathcal{F}}{\partial \rho^2} > 0 \ , \tag{22}$$

$$\frac{\partial^2 \mathcal{F}}{\partial \rho^2}\frac{\partial^2 \mathcal{F}}{\partial \rho_3^2} > \left(\frac{\partial^2 \mathcal{F}}{\partial \rho_3 \partial \rho}\right)^2 \ . \tag{23}$$

To make the physical content of these conditions more transparent, it is convenient to introduce the proton fraction $y$ defined by

$$y \equiv \frac{N_p}{N_p + N_n} = \frac{\rho + \rho_3}{2\rho} \tag{24}$$

and to rewrite the free-energy density as

$$\mathcal{F}(T, \rho, \rho_3) = \mathcal{F}(T, \rho, y) \ . \tag{25}$$

It is straightforward to show that the conditions (22) and (23) are equivalent to

$$\rho \left(\frac{\partial p}{\partial \rho}\right)_{T,y} = \rho^2 \left(\frac{\partial^2 \mathcal{F}}{\partial \rho^2}\right)_{T,y} > 0 \ , \tag{26}$$

$$\left(\frac{\partial \mu_p}{\partial y}\right)_{T,p} > 0 \quad \text{or} \quad \left(\frac{\partial \mu_n}{\partial y}\right)_{T,p} < 0 \ , \tag{27}$$

where we have introduced chemical potentials for protons and neutrons defined by

$$\mu_p = \mu + \mu_3 \quad , \quad \mu_n = \mu - \mu_3 \ . \tag{28}$$

The first inequality is the familiar requirement that the isothermal compressibility is positive, that is, that the system is mechanically stable. The second condition reflects the special character of the binary system. It expresses "diffusive stability", which guarantees that energy is required to change the concentration in a stable system while holding the remaining variables (pressure and temperature) fixed [42].

## III. RELATIVISTIC MEAN-FIELD EQUATION OF STATE

To describe the nuclear equation of state, we use a relativistic mean-field model containing nucleons, neutral scalar and vector fields, and the isovector $\rho$ meson field [22,23]. The neutral meson fields are self-interacting, including terms through fourth order in the fields, as proposed by Bodmer and Price [28,30]. (For simplicity, we omit couplings between the scalar and vector fields.) The $\rho$ meson is introduced in a minimal fashion, as discussed in the Introduction. This model can reproduce the observed properties of nuclear matter, and recent calculations show that it also gives an accurate description of the bulk and single-particle properties of nuclei throughout the periodic table [34].

The lagrangian density for this model can be written as

$$\begin{aligned}\mathcal{L} = &\overline{\psi}[\gamma_\mu(i\partial^\mu - g_{\rm v}V^\mu - \frac{1}{2}\,g_\rho\boldsymbol{\tau}\cdot\mathbf{b}^\mu) - (M - g_{\rm s}\phi)]\psi \\ &+ \frac{1}{2}\,(\partial_\mu\phi\partial^\mu\phi - m_{\rm s}^2\phi^2) - \frac{1}{3!}\,\kappa\phi^3 - \frac{1}{4!}\,\lambda\phi^4 \\ &- \frac{1}{4}\,F_{\mu\nu}F^{\mu\nu} + \frac{1}{2}\,m_{\rm v}^2 V_\mu V^\mu + \frac{1}{4!}\,\zeta g_{\rm v}^4(V_\mu V^\mu)^2 \\ &- \frac{1}{4}\,\mathbf{B}_{\mu\nu}\cdot\mathbf{B}^{\mu\nu} + \frac{1}{2}\,m_\rho^2\mathbf{b}_\mu\cdot\mathbf{b}^\mu\ .\end{aligned} \tag{29}$$

The scalar, isoscalar-vector, and isovector-vector fields are denoted by $\phi$, $V^\mu$, and $\mathbf{b}^\mu$, respectively, and the vector meson field strengths are $F^{\mu\nu} = \partial^\mu V^\nu - \partial^\nu V^\mu$ and $\mathbf{B}^{\mu\nu} = \partial^\mu\mathbf{b}^\nu - \partial^\nu\mathbf{b}^\mu - g_\rho\mathbf{b}^\mu\times\mathbf{b}^\nu$. We work in natural units with $\hbar = c = k_{\rm Boltzmann} = 1$.

In the mean-field approximation, the pressure $p$ and the energy density $\mathcal{E}$ are easy to compute and can be written as [22,23]

$$\begin{aligned}p = &\frac{1}{3\pi^2}\Big[H_5(\nu_p, M^*) + H_5(\nu_n, M^*)\Big] + \frac{m_{\rm v}^2}{2g_{\rm v}^2}\,W^2 + \frac{\zeta}{24}\,W^4 + \frac{m_\rho^2}{2g_\rho^2}\,R^2 \\ &- \frac{m_{\rm s}^2}{2g_{\rm s}^2}\,\Phi^2 - \frac{\kappa}{6g_{\rm s}^3}\,\Phi^3 - \frac{\lambda}{24g_{\rm s}^4}\,\Phi^4\ ,\end{aligned} \tag{30}$$

$$\begin{aligned}\mathcal{E} = &\frac{1}{\pi^2}\,\Big[H_5(\nu_p, M^*) + H_5(\nu_n, M^*) + M^{*2}H_3(\nu_p, M^*) + M^{*2}H_3(\nu_n, M^*)\Big] \\ &+ W\rho - \frac{m_{\rm v}^2}{2g_{\rm v}^2}\,W^2 - \frac{\zeta}{24}\,W^4 + \frac{1}{2}\,g_\rho R\rho_3 - \frac{m_\rho^2}{2g_\rho^2}\,R^2 \\ &+ \frac{m_{\rm s}^2}{2g_{\rm s}^2}\,\Phi^2 + \frac{\kappa}{6g_{\rm s}^3}\,\Phi^3 + \frac{\lambda}{24g_{\rm s}^4}\,\Phi^4\ ,\end{aligned} \tag{31}$$

with the conserved baryon density

$$\rho = \frac{1}{\pi^2}\Big[G_3(\nu_p, M^*) + G_3(\nu_n, M^*)\Big] \tag{32}$$

and isospin density

$$\rho_3 = \frac{1}{\pi^2}\Big[G_3(\nu_p, M^*) - G_3(\nu_n, M^*)\Big]\ . \tag{33}$$

Here, following Bodmer [30], we define the scaled meson fields $\Phi \equiv g_{\rm s}\phi$, $W \equiv g_{\rm v}V_0$, and $R \equiv g_\rho b_0$, with $b_0$ the timelike, neutral part of the $\rho$ meson field.

The baryon effective mass and effective chemical potentials are defined in terms of the meson mean fields as

$$M^* \equiv M - \Phi \ , \tag{34}$$

$$\nu_p \equiv \mu_p - W - \frac{1}{2} R \ , \tag{35}$$

$$\nu_n \equiv \mu_n - W + \frac{1}{2} R \ . \tag{36}$$

We also define the required integrals over the thermal distribution functions as

$$G_n(\mu, M) \equiv \int_0^\infty k^{n-1} dk \left( \frac{1}{1 + {\rm e}^{\beta[E(k,M)-\mu]}} - \frac{1}{1 + {\rm e}^{\beta[E(k,M)+\mu]}} \right) \ , \tag{37}$$

$$H_n(\mu, M) \equiv \int_0^\infty \frac{k^{n-1} dk}{E(k,M)} \left( \frac{1}{1 + {\rm e}^{\beta[E(k,M)-\mu]}} + \frac{1}{1 + {\rm e}^{\beta[E(k,M)+\mu]}} \right) \ , \tag{38}$$

where $E(k, M) \equiv (k^2 + M^2)^{1/2}$, and $n > 0$ to ensure convergence is understood. Note the important sign differences in these two relations. These functions obey the useful recursion relations (valid for $n > 1$)

$$\frac{\partial G_{n+1}}{\partial \mu} = n H_{n+1} + (n-1) M^2 H_{n-1} \ , \tag{39}$$

$$\frac{\partial G_{n+1}}{\partial M} = -(n-1) M G_{n-1} \ , \tag{40}$$

$$\frac{\partial H_{n+1}}{\partial \mu} = (n-1) G_{n-1} \ , \tag{41}$$

$$\frac{\partial H_{n+1}}{\partial M} = -(n-1) M H_{n-1} \ . \tag{42}$$

Thermodynamic equilibrium requires that the thermodynamic potential $\Omega$ be stationary with respect to changes in the mean fields, which leads to the self-consistency equations

$$\frac{m_{\rm s}^2}{g_{\rm s}^2} \Phi + \frac{\kappa}{2 g_{\rm s}^3} \Phi^2 + \frac{\lambda}{6 g_{\rm s}^4} \Phi^3 = \rho_s \ , \tag{43}$$

$$W \left( 1 + \frac{g_{\rm v}^2}{m_{\rm v}^2} \frac{\zeta}{6} W^2 \right) = \frac{g_{\rm v}^2}{m_{\rm v}^2} \rho \ , \tag{44}$$

$$R = \frac{g_\rho^2}{2 m_\rho^2} \rho_3 \ , \tag{45}$$

where the scalar density is given by

$$\rho_s = \frac{M^*}{\pi^2} \Big[ H_3(\nu_p, M^*) + H_3(\nu_n, M^*) \Big] \ . \tag{46}$$

These equations allow the fields to be held fixed when computing thermodynamic quantities as derivatives of the thermodynamic potential. Moreover, they allow the energy density to be expressed as

$$\mathcal{E} = \frac{1}{\pi^2}\Big[H_5(\nu_p, M^*) + H_5(\nu_n, M^*) + M^{*2}H_3(\nu_p, M^*) + M^{*2}H_3(\nu_n, M^*)\Big] + \frac{m_{\rm v}^2}{2g_{\rm v}^2}W^2 + \frac{\zeta}{8}W^4 + \frac{g_\rho^2}{8m_\rho^2}\rho_3^2 + \frac{m_{\rm s}^2}{2g_{\rm s}^2}\Phi^2 + \frac{\kappa}{6g_{\rm s}^3}\Phi^3 + \frac{\lambda}{24g_{\rm s}^4}\Phi^4 \ , \tag{47}$$

where $\Phi$ and $W$ are understood to satisfy the equations given above.

Finally, we exhibit the corresponding expression for the entropy density, which follows from the Gibbs' relation:

$$\begin{aligned}\sigma &= (p + \mathcal{E} - \mu_p\rho_p - \mu_n\rho_n)/T \\ &= \frac{1}{3\pi^2 T}\Big[4H_5(\nu_p, M^*) + 4H_5(\nu_n, M^*) \\ &\qquad + 3M^{*2}\Big(H_3(\nu_p, M^*) + H_3(\nu_n, M^*)\Big) \\ &\qquad - 3\nu_p G_3(\nu_p, M^*) - 3\nu_n G_3(\nu_n, M^*)\Big] \ .\end{aligned} \tag{48}$$

Note that none of the previous relations involve thermal contributions from the mesons, since their masses are too large for these to be relevant. Indeed, at the temperatures of interest in this work, the antibaryon contributions are negligible as well.

To specify the parameters, we observe that models that successfully reproduce bulk and single-particle properties of finite nuclei share characteristic properties in infinite nuclear matter [32,44]. After taking the zero-temperature limit of the preceding results, one can obtain an explicit set of transcendental equations that determines the parameters for the desired choice of nuclear matter properties. (We will not exhibit these equations here; see Refs. [30,34].) The parameters so obtained are listed in Table I.

TABLE I. Mean-Field Parameters

| $C_{\rm s}^2$ | $C_{\rm v}^2$ | $C_\rho^2$ | $\kappa/M$ | $\lambda$ | $\zeta$ |
|---|---|---|---|---|---|
| 374.77 | 260.57 | 106.91 | 3.0809 | 8.106 | 0.02364 |

TABLE II. Nuclear Matter Properties

Equilibrium Properties:

| $k_{\rm F}$ | $\rho$ | $M^*/M$ | $e_0$ | $K_V^{-1}$ | $a_4$ |
|---|---|---|---|---|---|
| $1.30\,{\rm fm}^{-1}$ | $0.1484\,{\rm fm}^{-3}$ | 0.60 | $-15.75$ MeV | 250 MeV | 35 MeV |

Critical Values:

| $T_c$ | $\rho_c$ | $M_c^*/M$ | $p_c$ |
|---|---|---|---|
| 14.40 MeV | $0.04661\,{\rm fm}^{-3}$ | 0.8543 | $0.2010\,{\rm MeV/fm}^3$ |

Note that the nucleon and vector meson masses are chosen to take their empirical values ($M = 939\,\mathrm{MeV}$, $m_{\mathrm{v}} = m_\omega = 783\,\mathrm{MeV}$, $m_\rho = 770\,\mathrm{MeV}$), and only the ratios of couplings to masses (denoted by $C_i^2 \equiv g_i^2 M^2/m_i^2$) are needed in infinite matter. The resulting properties of nuclear matter, as well as the properties at the critical point in symmetric matter, are given in Table II. To generate acceptable bulk nuclear properties, it is important to accurately reproduce the nuclear matter equilibrium density, energy/nucleon $e_0$, baryon effective mass $M^*$, compressibility $1/K_V$ [45], and bulk symmetry energy $a_4$. The first three of these are tightly constrained [32], whereas the latter two are not. We will begin by studying warm nuclear matter for the values of $K_V$ and $a_4$ given in Table II and later examine the sensitivity to reasonable variations in these values. The observant reader will notice that the model has six free parameters (in nuclear matter) that are determined by only five constraints. Thus there are actually an infinite number of parameter sets that will reproduce the equilibrium properties listed in Table II. These sets differ in the way the nonlinear meson interactions are split between the scalar terms ($\kappa, \lambda$) and the vector term ($\zeta$). To arrive at the parameter values in Table I, Dirac–Hartree calculations of finite nuclei were also performed in this model [34], and the parameters were tuned to give optimal bulk- and surface-energy systematics.[1] This allows us to proceed with the most realistic mean-field nuclear equation of state possible.

We close this section with some remarks concerning the numerical procedures. Although the solution of the self-consistent equations (43)–(45) is in principle straightforward, the analysis becomes involved due to multiple roots. For example, at low temperatures, a given set $\{\nu_p, \nu_n\}$ leads to either one or three solutions for the scalar field in Eq. (43). We resolve this problem using the crucial fact that any quantity can be uniquely and continuously parametrized in terms of the effective mass $M^* = M - \Phi$. To give a concrete example, consider the equation of state at constant pressure and proton fraction. In this case we need to solve three equations, namely, Eq. (30) with a given value of $p$ on the left-hand side, an equation that fixes the proton concentration $y$,

$$y = \frac{G_3(\nu_p, M^*)}{G_3(\nu_p, M^*) + G_3(\nu_n, M^*)} \; ,$$

and Eq. (43). This set is solved for a given value of the effective mass leading to a *unique* root of the form $\{\nu_p(M^*), \nu_n(M^*), T(M^*)\}$, which can in turn be used to evaluate all the remaining quantities of interest, *e.g.*, the density and entropy. Once the solutions have been obtained for a given $M^*$, we can proceed to map out all the desired variables by making small incremental changes in this parameter.

The main ingredient in our thermodynamic treatment is the binodal surface, namely, the collection of points in parameter space that satisfy the Gibbs' conditions (14) and (15). Numerically this surface is most easily parametrized in terms of the pressure, so that the equations we solve simultaneously are

$$p = p(\nu_p', \nu_n', M^{*\prime}) \; , \tag{49}$$

$$p = p(\nu_p'', \nu_n'', M^{*\prime\prime}) \; , \tag{50}$$

---

[1]The authors are grateful to R. J. Furnstahl for his assistance in obtaining the parameter sets.

$$\mu_{n,p}(\nu_p', \nu_n', M^{*\prime}) = \mu_{n,p}(\nu_p'', \nu_n'', M^{*\prime\prime}) \ , \tag{51}$$

$$\frac{m_{\rm s}^2}{g_{\rm s}^2}\,\Phi' + \frac{\kappa}{2g_{\rm s}^3}\,\Phi'^2 + \frac{\lambda}{6g_{\rm s}^4}\,\Phi'^3 = \rho_s(\nu_p', \nu_n', M^{*\prime}) \ , \tag{52}$$

$$\frac{m_{\rm s}^2}{g_{\rm s}^2}\,\Phi'' + \frac{\kappa}{2g_{\rm s}^3}\,\Phi''^2 + \frac{\lambda}{6g_{\rm s}^4}\,\Phi''^3 = \rho_s(\nu_p'', \nu_n'', M^{*\prime\prime}) \ , \tag{53}$$

for given values of $p$ and $T$. This procedure works well except near the critical points, where the two solutions coalesce. Correspondingly, results in this region must be obtained by interpolation, after one determines the location of the critical points using Eq. (54), below.

## IV. APPLICATION TO ASYMMETRIC NUCLEAR MATTER

Phase transitions in binary systems are more complex than in one-component systems. In the case of nuclear matter, as discussed in Sec. II, the global instability boundary forms a two-dimensional surface in $(T, p, y)$ space, enclosing the region where either mechanical instability [Eq. (26)] or diffusive instability [Eq. (27)] occurs.

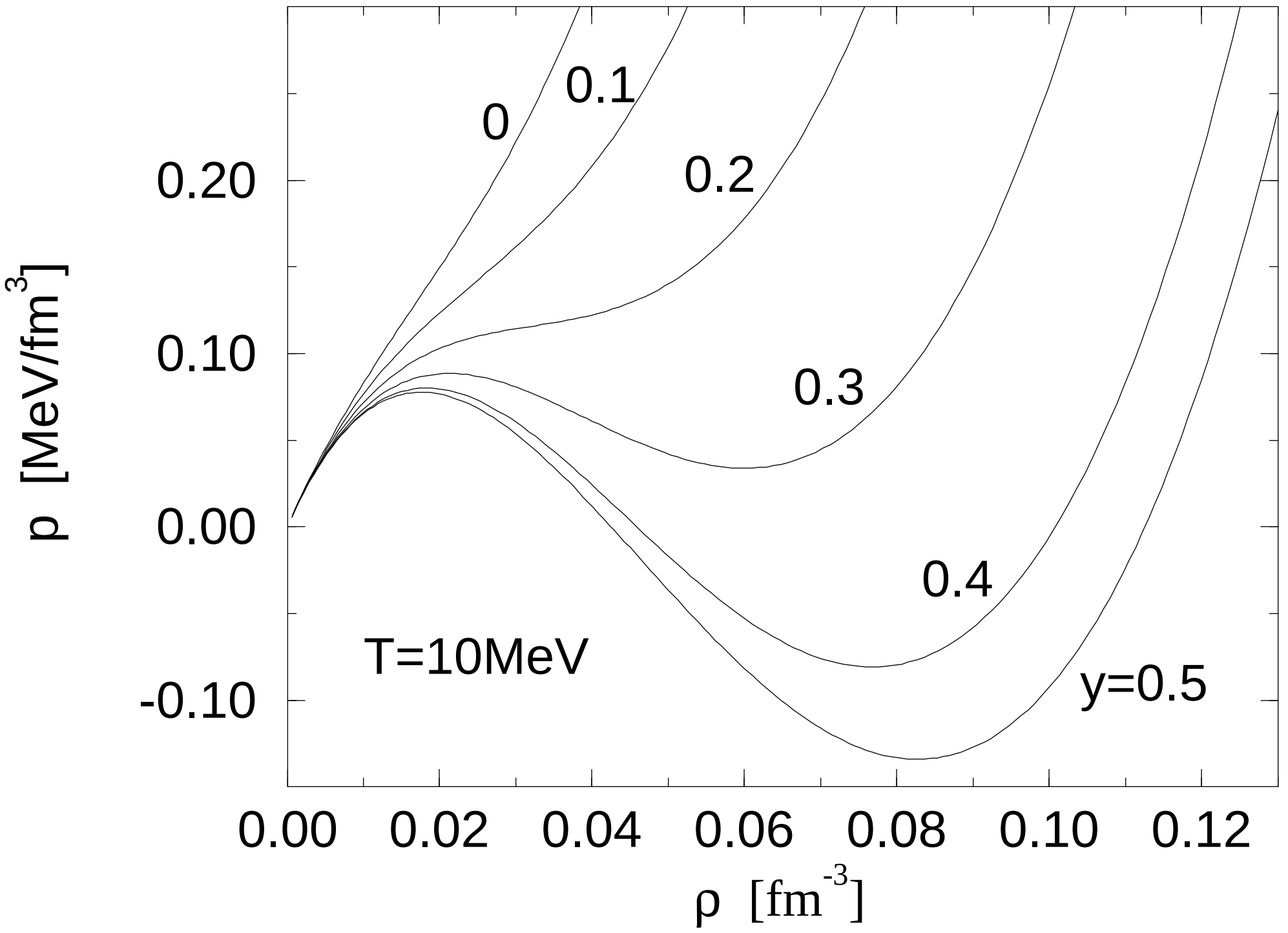

FIG. 5. Pressure as a function of baryon density at fixed temperature for various proton fractions $y$.

To be more specific, Fig. 5 shows the pressure as function of the baryon density at a fixed temperature but for different proton fractions, using the parameter set in Table I. For small $y$, and in particular, for pure neutron matter ($y = 0$), the pressure increases monotonically, so the matter is stable at all densities. In contrast, for $y \gtrsim 0.2$, the compressibility becomes negative, indicating a mechanical instability. The full complexity of the binary system is indicated in Fig. 6, where chemical potential isobars for neutrons and protons are shown as a function of $y$ at fixed temperature. Above a certain critical pressure $p_c$, the matter is stable, but for $p < p_c$, the second condition (27) is violated, and the system becomes chemically unstable. The critical isobar $p_c$ is determined by the existence of an inflection point:

$$\left(\frac{\partial \mu_p}{\partial y}\right)_{T,p} = \left(\frac{\partial^2 \mu_p}{\partial y^2}\right)_{T,p} = 0 \ . \tag{54}$$

This isobar marks the upper boundary of instability with respect to the pressure and defines a critical point $(p_c, y_c)$ for a given temperature [42].

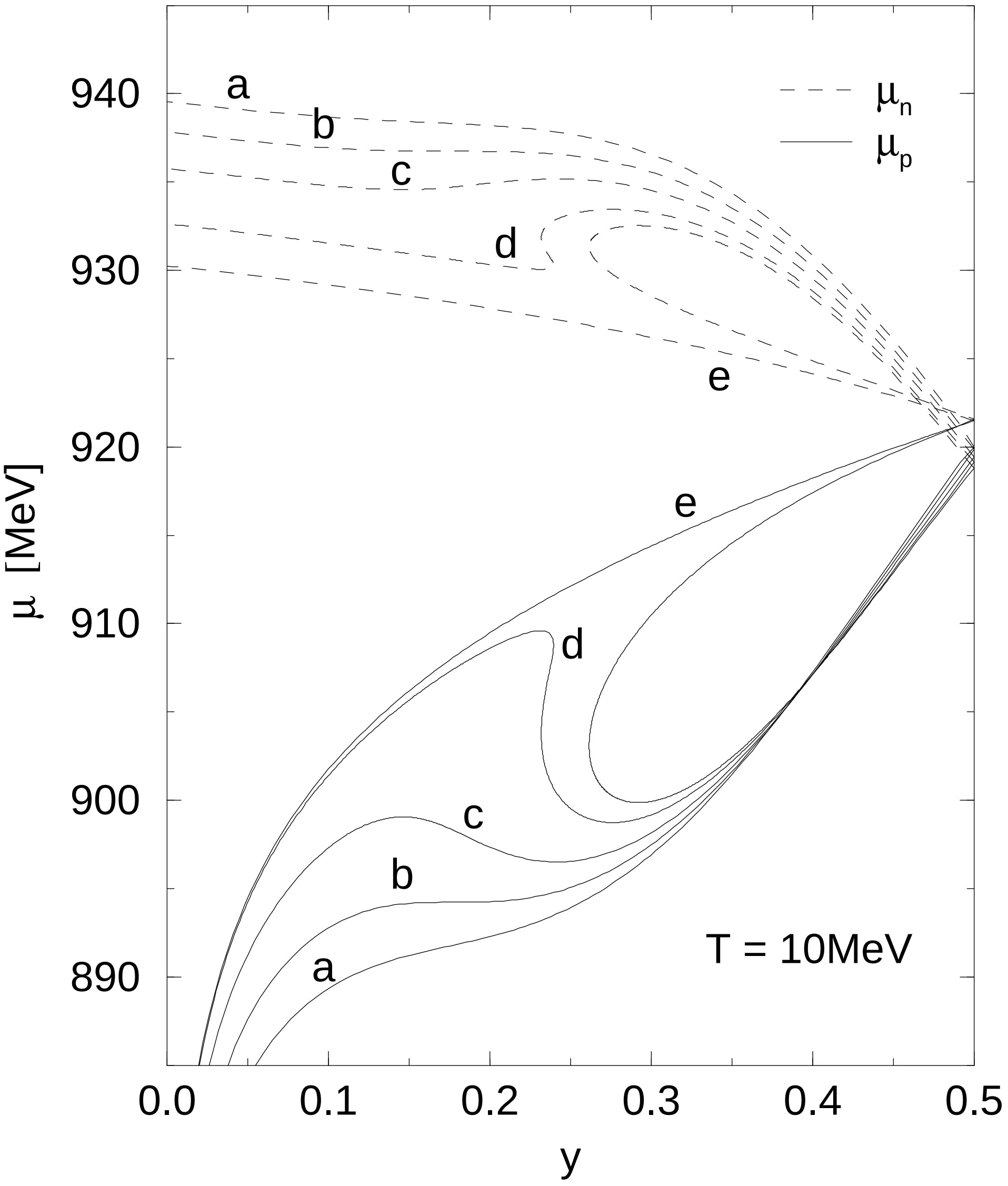

FIG. 6. Chemical potential isobars at fixed temperature as a function of $y$. The curves labeled $a$ through $e$ have pressures $p = 0.25, 0.198, 0.15, 0.10, 0.075\,\mathrm{MeV/fm^3}$, respectively. The curves labeled $b$ are at the critical pressure $p_c = 0.198\,\mathrm{MeV/fm^3}$.

The Gibbs' conditions (14) and (15) for phase equilibrium demand equal pressure and chemical potentials for two phases with different concentrations. Thus the two desired solutions form the edges of a rectangle and can be found by means of the geometrical construction shown in Fig. 7 [19]. The collection of all such pairs $y_1(T,p)$ and $y_2(T,p)$ form the binodal surface. For a given temperature, the two-phase region is limited from below by the pressure at equal concentrations $p_{\rm eq}$, which corresponds to symmetric nuclear matter ($y = 0.5$). Correspondingly, the rectangle in Fig. 7 shrinks to a point at this particular pressure, and the system becomes stable again for $p < p_{\rm eq}$. We did not find any cases of three-phase coexistence, which would require two rectangles with a common side.

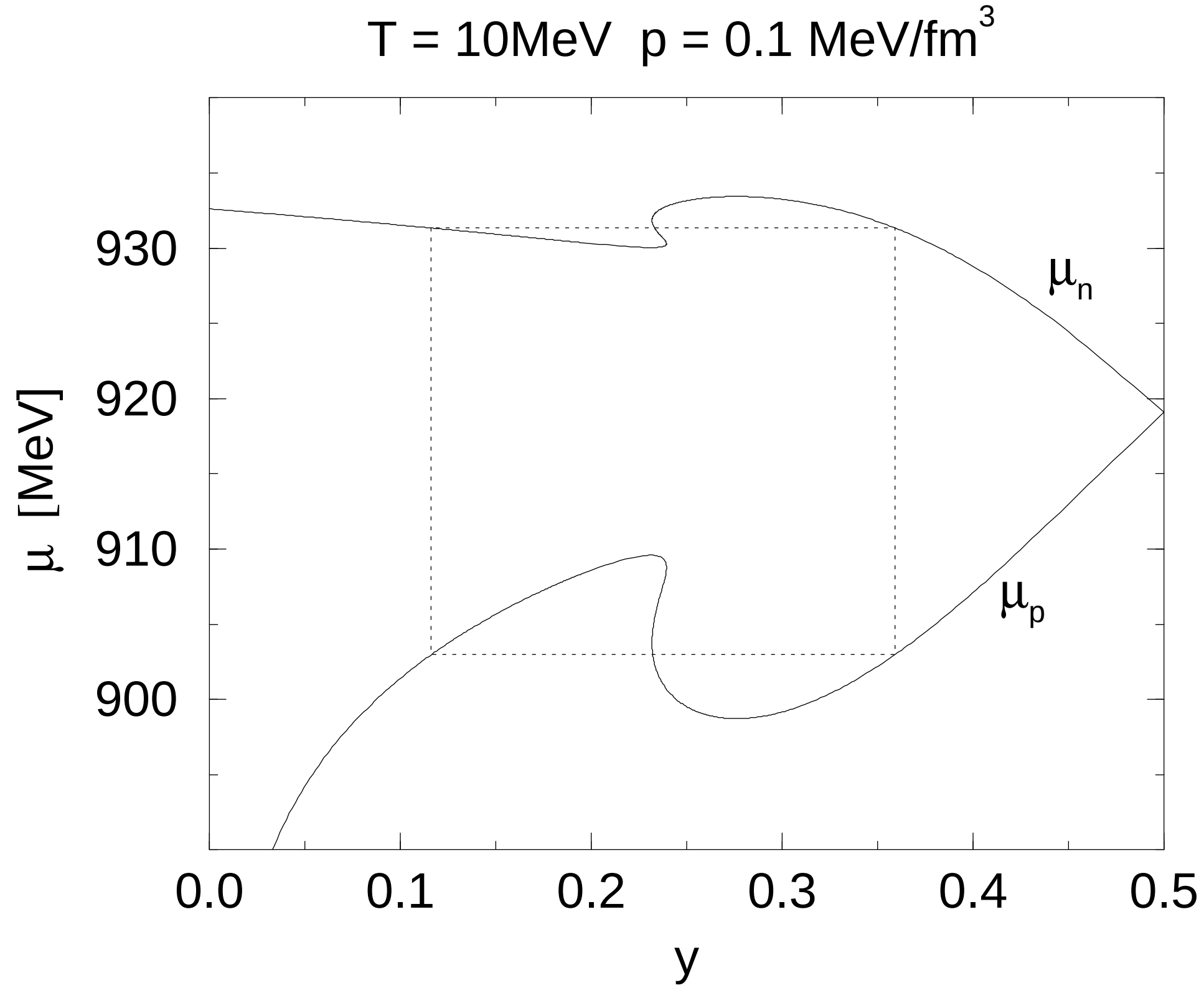

FIG. 7. Geometrical construction used to obtain the proton fractions and chemical potentials in the two coexisting phases at fixed $T$ and $p$.

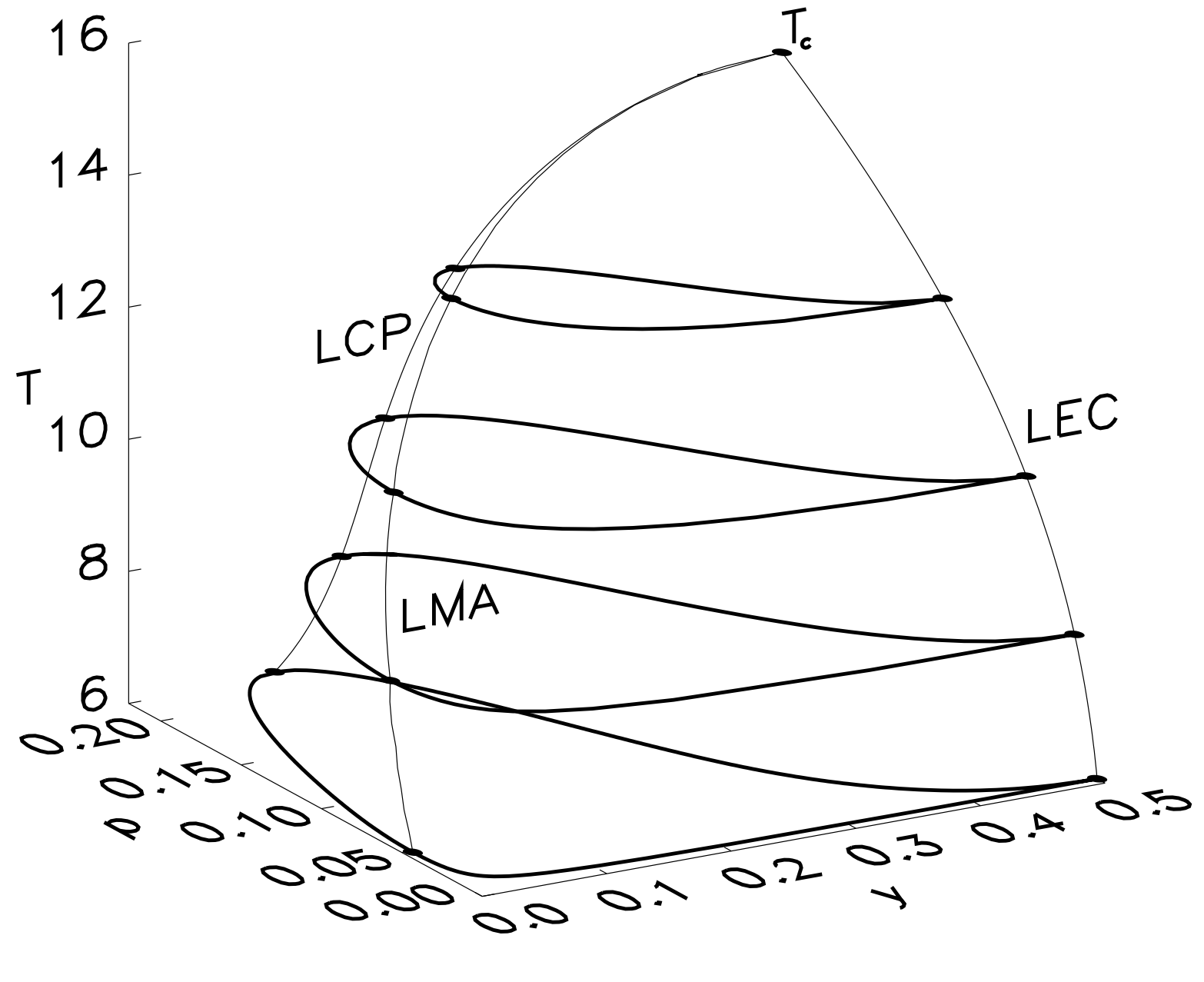

FIG. 8. The binodal surface indicating the two-dimensional phase-coexistence boundary is shown in $(p, T, y)$ space. The critical temperature $T_c(y = 0.5)$, the line of equal concentrations (LEC), the line of critical points (LCP), and the line of maximal asymmetry (LMA) are indicated. $T$ is in MeV, $p$ is in MeV/fm$^3$, and $y$ is dimensionless.

The phase-separation boundary, or binodal surface, obtained from the preceding geometrical constructions [see Eqs. (49) through (53)] is indicated in Fig. 8. (The shape of this surface has been previously described as a "filet mignon" [19].) Several slices at constant $T$ are indicated, and one observes that the enclosed area in these sections decreases with increasing temperature until it vanishes at the critical point $T_c$ of symmetric nuclear matter. This is also the point at which the lines LCP and LMA meet at $y = 0.5$. The line of critical points (LCP) is determined by the solutions of Eq. (54), and it can be parametrized uniquely by just one of the coordinates $(T, p, y)$. The LCP begins at the critical point of symmetric nuclear matter ($y = 0.5, T = T_c$) and ends at $y = y_0 = 0.057$ at $T = 0$. Note also that for a given temperature, the critical point determines the maximum pressure in the two phase region.

In addition to the LCP, we have also indicated the points on the binodal surface with the maximal asymmetry (LMA), or minimal proton fraction $y_{\rm min}$, at each temperature.[2] Any system with $y < y_{\rm min}$ is external to the two-phase region and is therefore stable. Moreover, since $y_{\rm min}$ is a monotonically increasing function of the temperature, the LMA determines the maximum temperature $T_{\rm max}$ of phase separation at any given $y$ [19]. The LMA also begins at the critical point of symmetric nuclear matter and descends to a small, positive value of $y_{\rm min}$ at $T = 0$. This implies that pure neutron matter is stable at all temperatures in this model.

Configurations that separate into two phases each having equal numbers of neutrons and protons form the line of equal concentration (LEC). In these cases, the binodal section at any $T$ degenerates into a point, at which the relation [41,43]

$$\left(\frac{\partial p}{\partial y}\right)_T = 0 \qquad (55)$$

is satisfied. (This is an example of the Gibbs–Konowalow rule [43].) The LEC coincides with the projection $y = 0.5$ in our model, which implies that only symmetric nuclear matter will separate into phases with equal concentration [19]. Since symmetric matter behaves as a one-component system,[3] one observes how the phase separation simplifies in this case: in an isothermal compression, the system evolves until it encounters the binodal (which is just the LEC) and then remains there until the transition is complete. In contrast, for $y \neq 0.5$, the system encounters a two-dimensional section of the binodal surface. Since the energy contains a term proportional to $\rho_3^2$ [see Eq. (47)], it is energetically favorable for asymmetric matter to separate into a liquid phase that is less asymmetric and a gas phase that is more asymmetric, rather than into two phases with equal concentration. This leads to more complex phase separations, as we discuss shortly.

[2]As with all results in this section, our analysis is symmetric with respect to protons and neutrons, so that only the physically useful regime $0 \le y \le 0.5$ need be considered.

[3]Strictly speaking, symmetric nuclear matter is an *azeotrope*. During the liquid–gas phase separation, all equilibrium configurations have two phases with the same composition ($y = 0.5$). This type of phase equilibrium is called "indifferent equilibrium" [43]. Since the liquid–gas phase equilibrium in a one-component system is also indifferent, symmetric nuclear matter behaves as a one-component system.

For a given temperature, the binodal section is divided into two branches by the critical point and the point of equal concentration. One branch describes the system in a high-density (liquid) phase, while the other branch describes the low-density (gas) phase. These two branches contain the beginning and ending configurations of the phase transition.

We now return to the behavior of the matter under isothermal compression, to illustrate the different phase-separation scenarios.

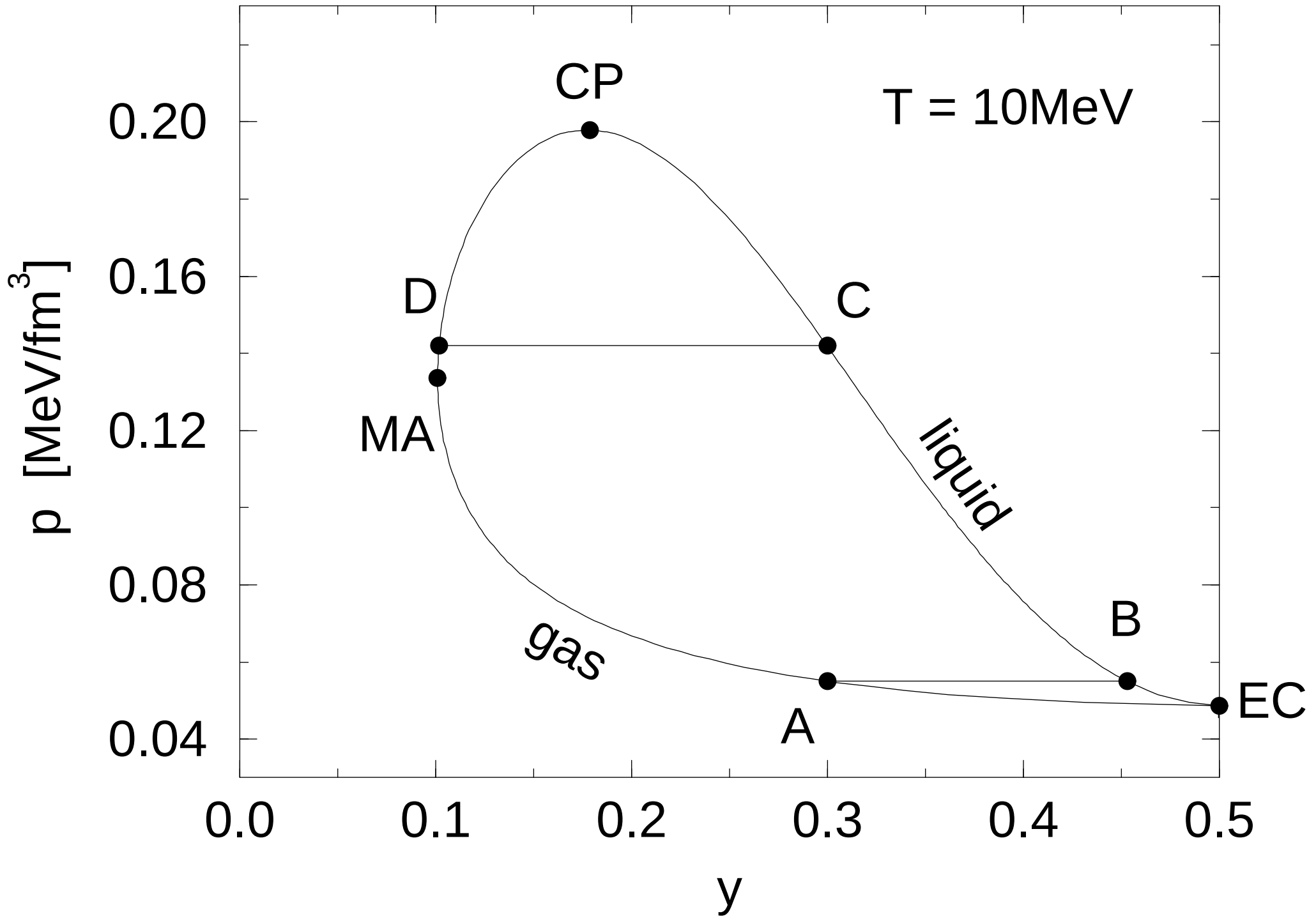

FIG. 9. Binodal section at $T = 10\,\mathrm{MeV}$. The points $A$ through $D$ denote phases participating in a normal phase transition. The critical point (CP) and the points of equal concentration (EC) and maximal asymmetry (MA) are also indicated.

Consider the situation in Fig. 9, which shows a section through the binodal surface at $T = 10\,\mathrm{MeV}$. The critical point CP, the point of maximal asymmetry MA, and the point of equal concentration EC are indicated, and the validity of Eq. (55) is apparent. Assume that the system is initially prepared in the low-density (gas) phase with proton fraction $y = 0.3$. During the compression, the two-phase region is encountered at the point $A$, and now a (liquid) phase with a higher density begins to emerge. The geometrical construction described above determines the density and the proton fraction $y_B$ of this new phase, which occurs at the point labeled $B$. As the system is compressed, the *total* proton fraction $y$ remains fixed, as dictated by the conservation laws, but the gas phase evolves from $A$ to $D$, while the liquid phase evolves from $B$ to $C$. At the point $C$, the system leaves the region of instability. The original (gas) phase is about to disappear, and it exists in an infinitesimal volume with a density and proton fraction $y_D$ corresponding to the point $D$.

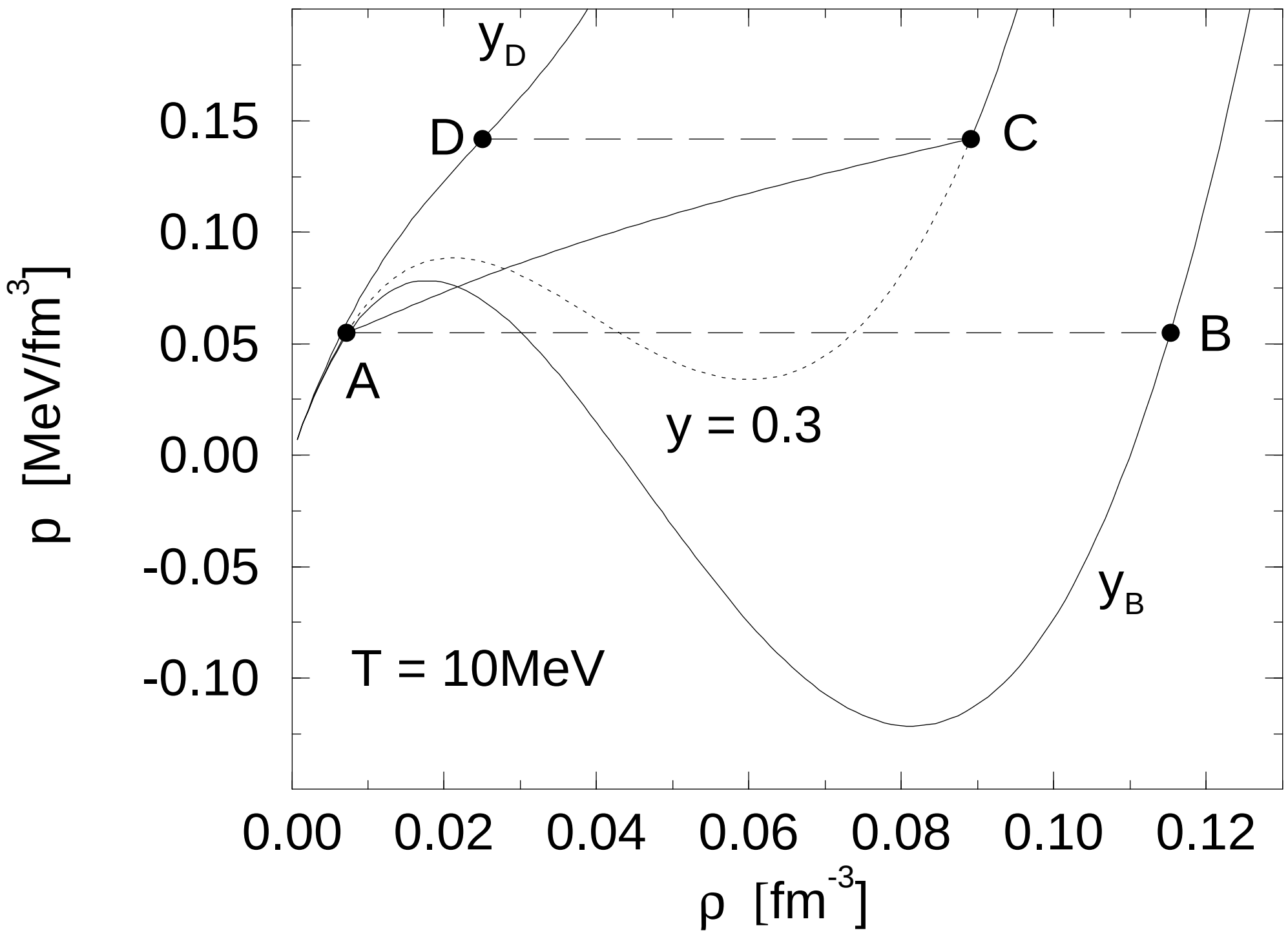

FIG. 10. Isotherms for a normal phase transition at $T = 10\,\mathrm{MeV}$ and initial condition $y = 0.3$. The Maxwell construction produces the curve $AC$. Note that $y_D \neq 0.3 \neq y_B$.

To evolve the system between configurations $A$ and $C$, we must solve the equations

$$\rho = (1-\lambda)\rho' + \lambda\rho'' \;, \tag{56}$$

$$\rho_3 = (1-\lambda)\rho_3' + \lambda\rho_3'' \;, \tag{57}$$

according to (18), for densities that lie on the binodal surface, and for given values of $\rho$ and $\rho_3 = (2y-1)\rho$. The result is the generalized Maxwell construction in the binary system. The corresponding isotherms are drawn in Fig. 10. The dotted line between $A$ and $C$ is the unphysical course of the pressure at the fixed total proton fraction, and the nearly straight line connecting $A$ and $C$ is the interpolation due to the Maxwell construction, which corresponds to the stable configuration at each intermediate density. As mentioned earlier, the compression in the two-phase region is nonzero because of the pressure change forced by the conservation laws, in contrast to the behavior found in a single-component system. The volume fraction $\lambda$ starts with $\lambda = 0$ at $A$ and runs through the whole interval $[0,1]$, ending with $\lambda = 1$ at $C$. Since the points $A$ and $C$ lie on different branches of the binodal surface (as defined above), the matter has undergone a phase transition from a gas to a liquid phase.

Interestingly enough, the geometry of the binodal surface offers a second possibility. In the previous example, there is a transition between the two branches of the binodal surface because the value of $y$ in the original phase is larger than $y_c$. For $y < y_c$, however, the system enters and leaves the two-phase region on the same branch, so the system remains in the same phase.

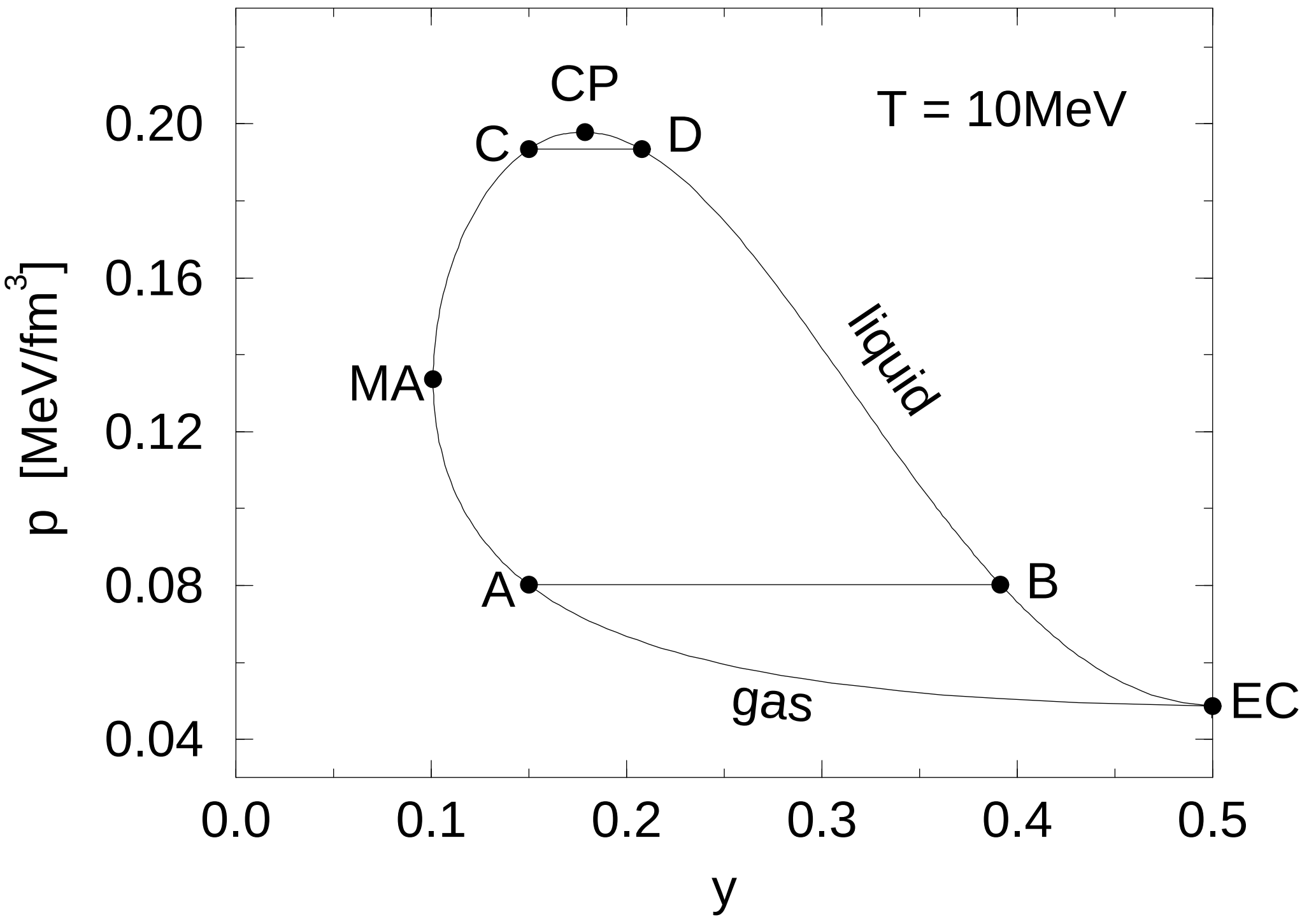

FIG. 11. Binodal section at $T = 10\,\mathrm{MeV}$. The points $A$ through $D$ denote phases participating in a retrograde phase transition.

This situation is depicted in Fig. 11. We consider the $T = 10\,\mathrm{MeV}$ isotherm and prepare the system with $y = 0.15 < y_c$. The system becomes unstable at the point $A$, and as before, a liquid phase with a higher density begins to emerge at $B$. The system is compressed at fixed total $y$, with the liquid phase evolving from $B$ to $D$, and the gaseous phase, from $A$ to $C$. At $C$ the system crosses the binodal again, but this time on the same branch, that is, still in the original (gas) phase. The high-density (liquid) phase, now at $D$, vanishes at this stage. The Maxwell construction for the corresponding isotherms, which follows from the solution of Eqs. (56) and (57), is represented by the solid line connecting $A$ and $C$ in Fig. 12. Note that this new phenomenon is caused solely by a diffusive instability. The matter remains mechanically stable throughout the entire process, as indicated by the dotted curve, but it is energetically favorable to separate into two phases with different proton fractions. In contrast to the previous case, one also finds a different behavior for $\lambda$. The initial value $\lambda = 0$ at $A$ increases up to a maximal value $\lambda_{\mathrm{max}} < 1$ and then decreases to zero when $C$ is reached. Thus, although a second phase is present between $A$ and $C$, the system does not convert completely; on the contrary, the second phase vanishes steadily after having occupied a maximum volume fraction ($\lambda_{\mathrm{max}}$). This *retrograde condensation* is unique to the binary system and does not occur in one-component systems.

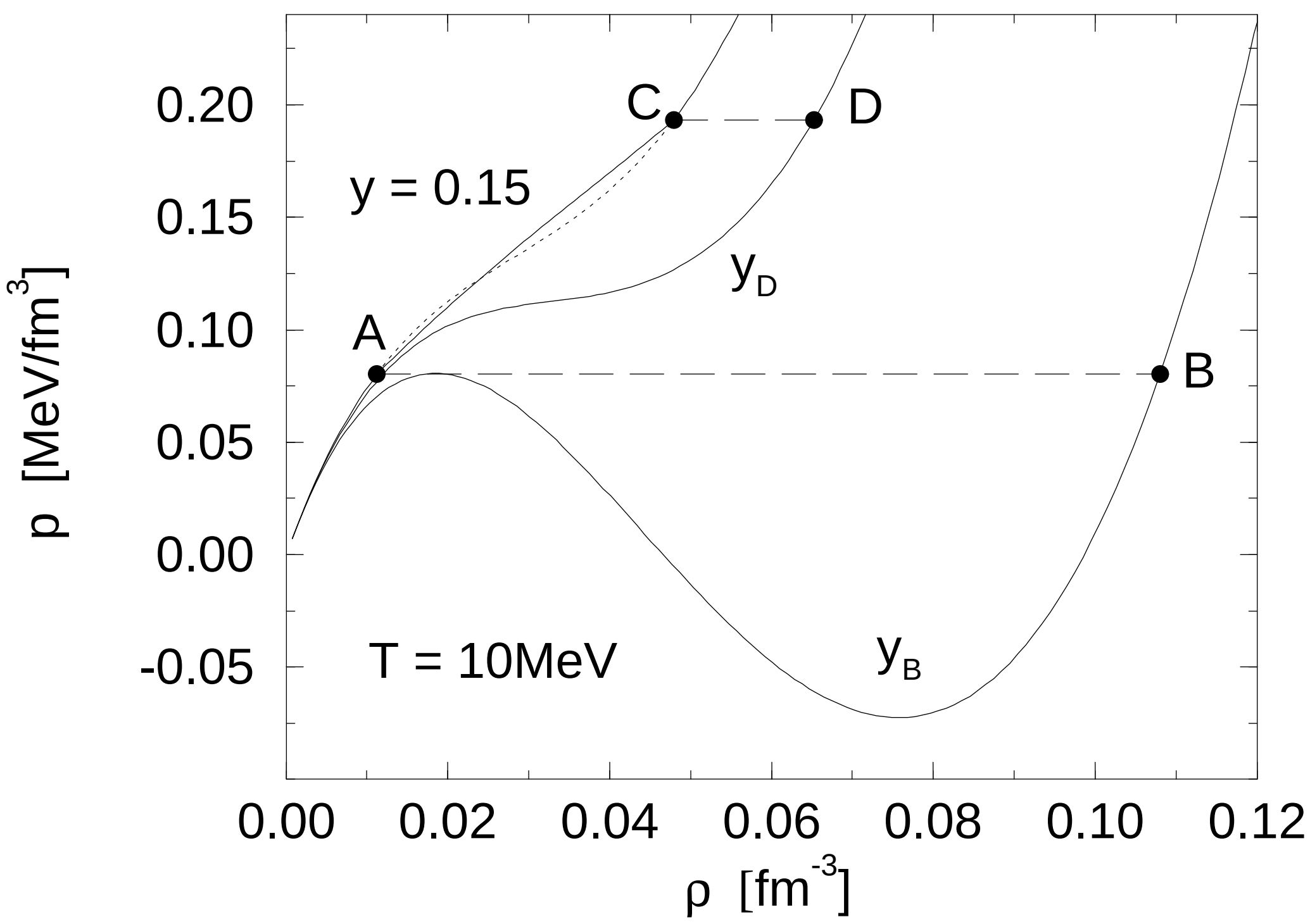

FIG. 12. Isotherms for a retrograde phase transition at $T = 10\,\mathrm{MeV}$ and initial condition $y = 0.15$. The Maxwell construction produces the curve $AC$.

A more complete picture of the different Maxwell constructions is shown in Fig. 13. In part (a), we consider different proton fractions at a fixed temperature. At $y = 0.5$ (symmetric matter), we obtain the familiar result with a constant vapor pressure, represented by a horizontal line. With increasing asymmetry (decreasing $y$), the compression increases in the two-phase region, as the vapor pressure is no longer constant. For $y_{\rm min} < y < y_c$, the matter is mechanically stable, and the system undergoes retrograde condensation. Finally, the system becomes completely stable for $y < y_{\rm min}$. Part (b) shows similar Maxwell constructions on different isotherms at fixed proton fraction $y = 0.4$. For a more thorough discussion of the variation of the densities in the two phases throughout the transition, see Ref. [19].

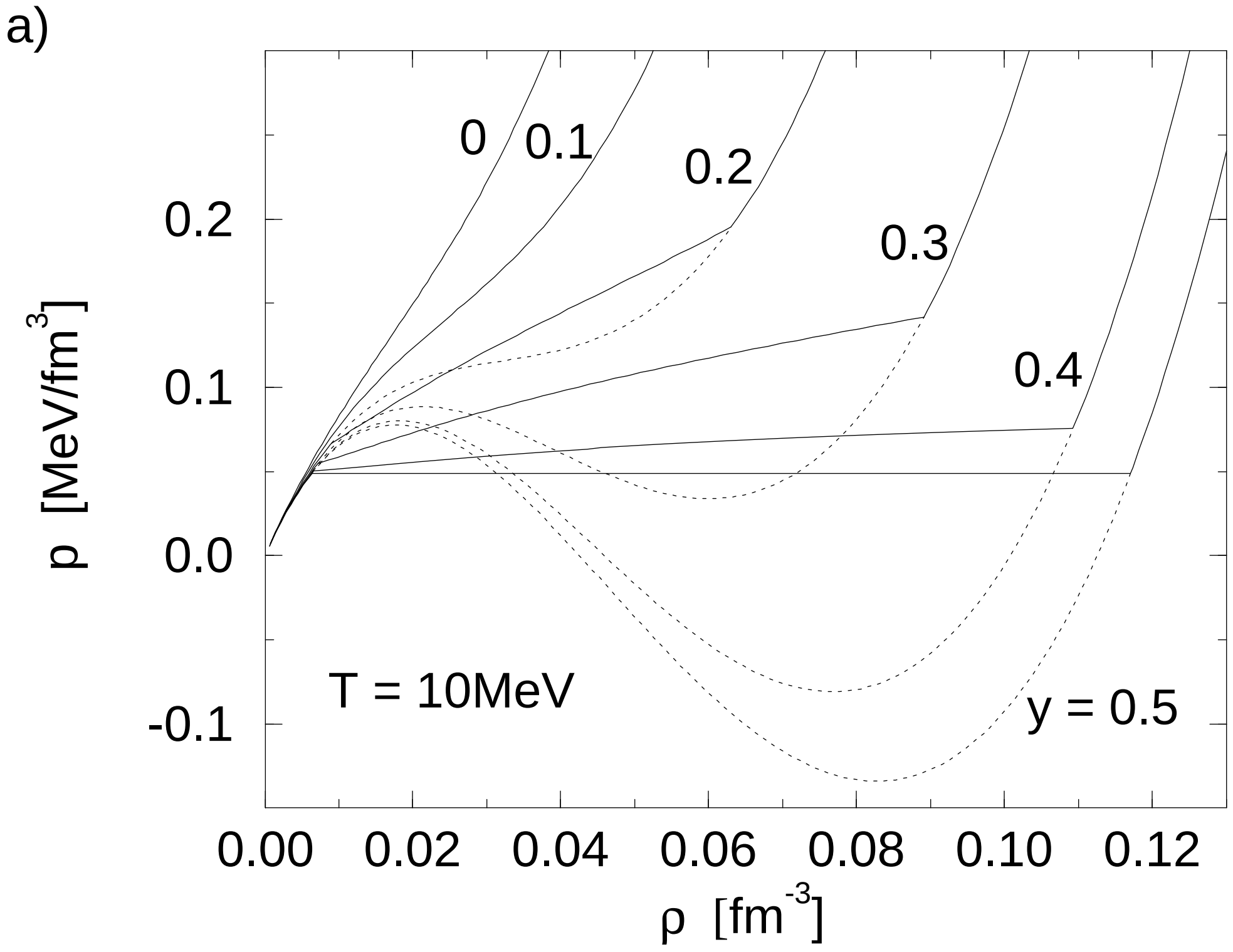

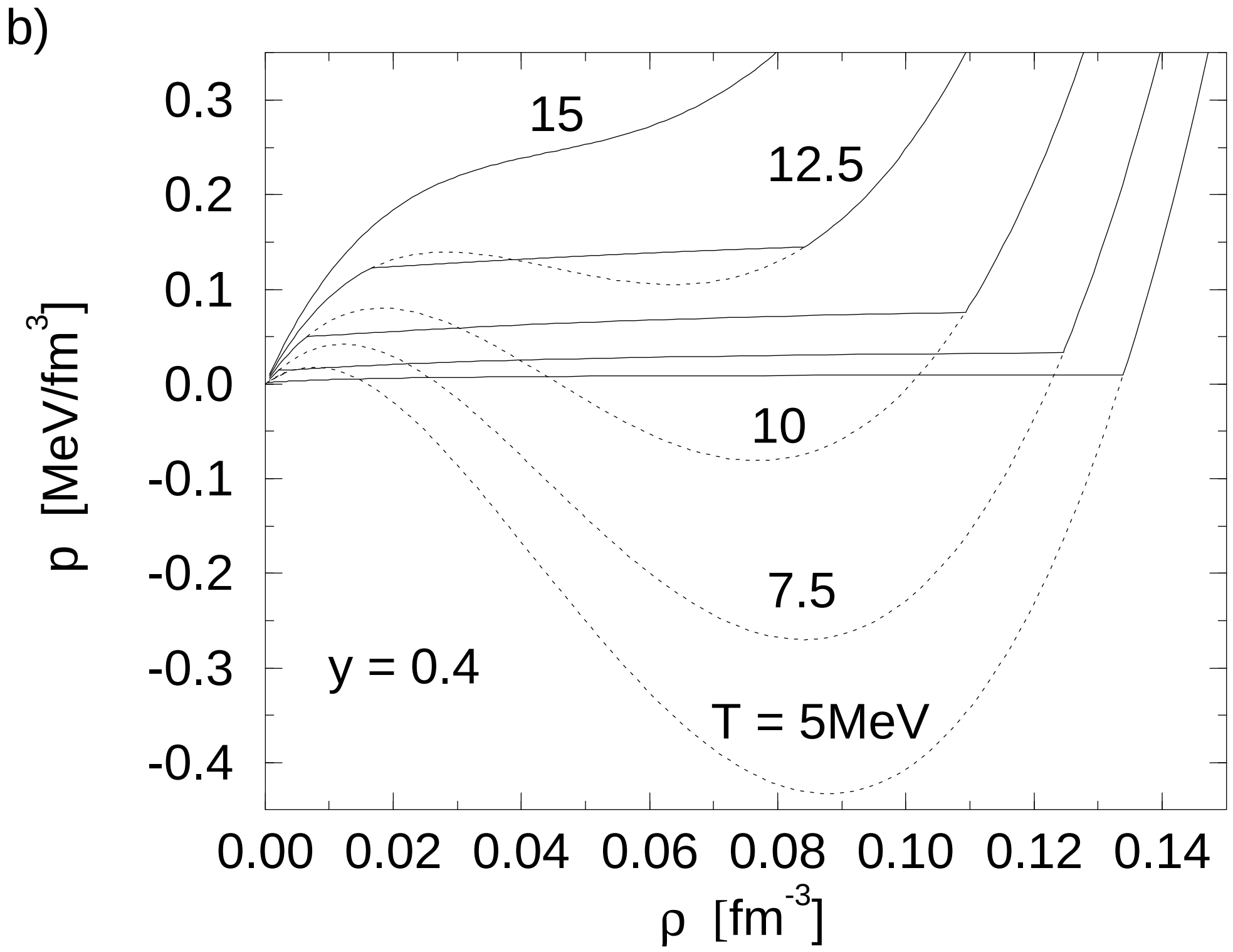

FIG. 13. (a) Maxwell constructions at fixed temperature for various proton fractions. (b) Maxwell constructions at fixed proton fraction for various temperatures.

To close this section, we consider the order of the liquid–gas phase transition. First-order phase transitions in single-component systems are characterized by discontinuities in certain physical quantities. This occurs because quantities like the density and the entropy are different in the two distinct phases, but they remain constant throughout the phase transition. This behavior does not occur in the binary system, because the constraints of charge conservation and Gibbs' criteria force the density (and pressure) in each individual phase to change throughout the transition. We might therefore expect the transition in the binary system to be "smoother". To make this point more precise, we consider the Gibbs free energy (or free enthalpy) per nucleon

$$\frac{G(T,p,N_p,N_n)}{N_p+N_n} \equiv g(T,p,y) = y\mu_p + (1-y)\mu_n \ , \tag{58}$$

which allows us to discuss isobaric processes as a function of temperature.

This quantity is shown for several different asymmetries in Fig. 14. As discussed in Sec. II, the Gibbs free energy must be computed according to

$$g(T,p,y) = (1-\bar{\lambda})g(T,p,y') + \bar{\lambda}g(T,p,y'') \tag{59}$$

in the transition region, where $\bar{\lambda}$ now specifies the number of particles in each phase:

$$\bar{\lambda} \equiv \frac{N_p''+N_n''}{N_p+N_n} \ , \tag{60}$$

and where $y'$ and $y''$ denote the different proton fractions. If one heats the symmetric system through the transition point, as indicated in Fig. 14a, it remains at constant temperature until the transition is completed, producing a striking kink in the free enthalpy curve.

a)

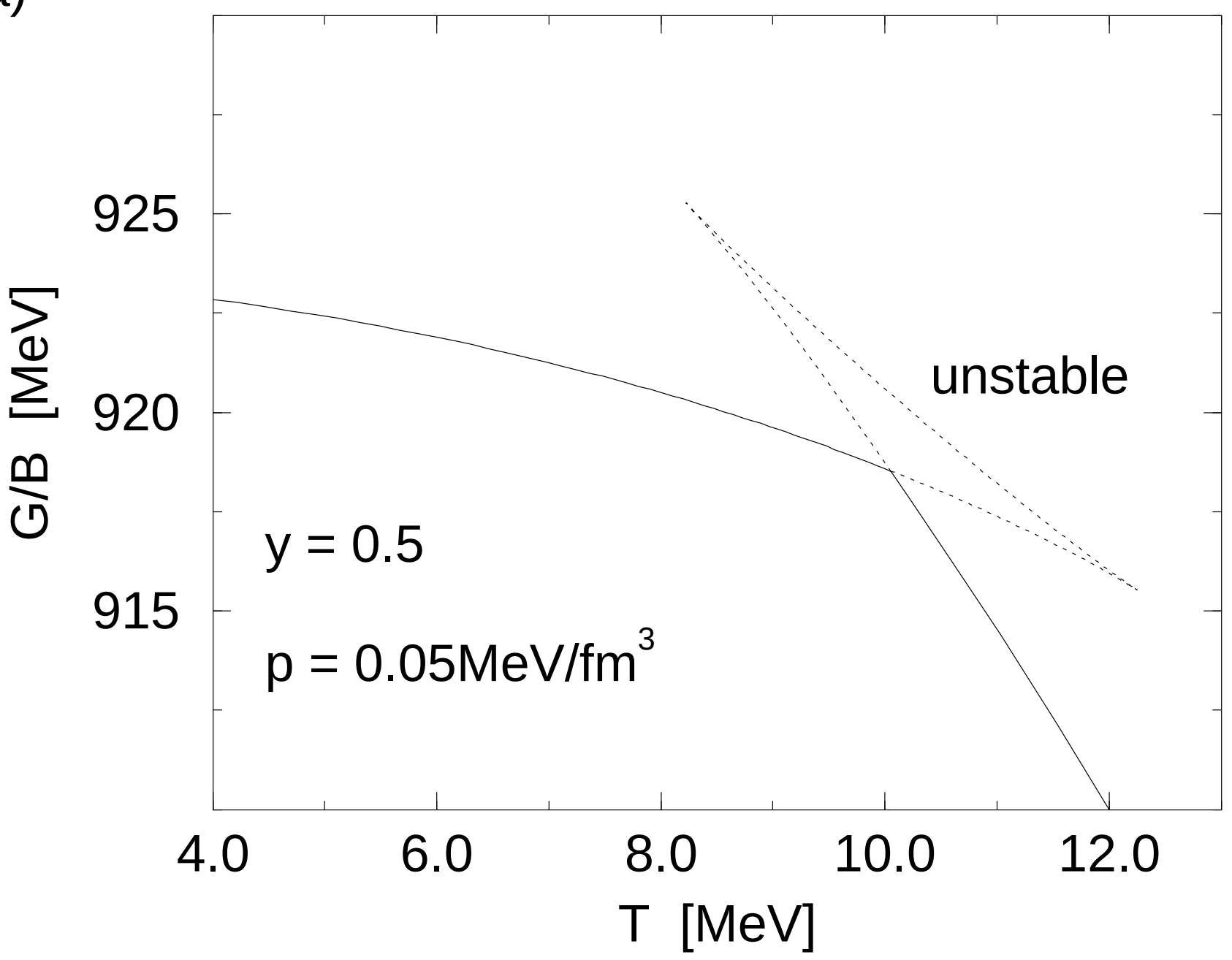

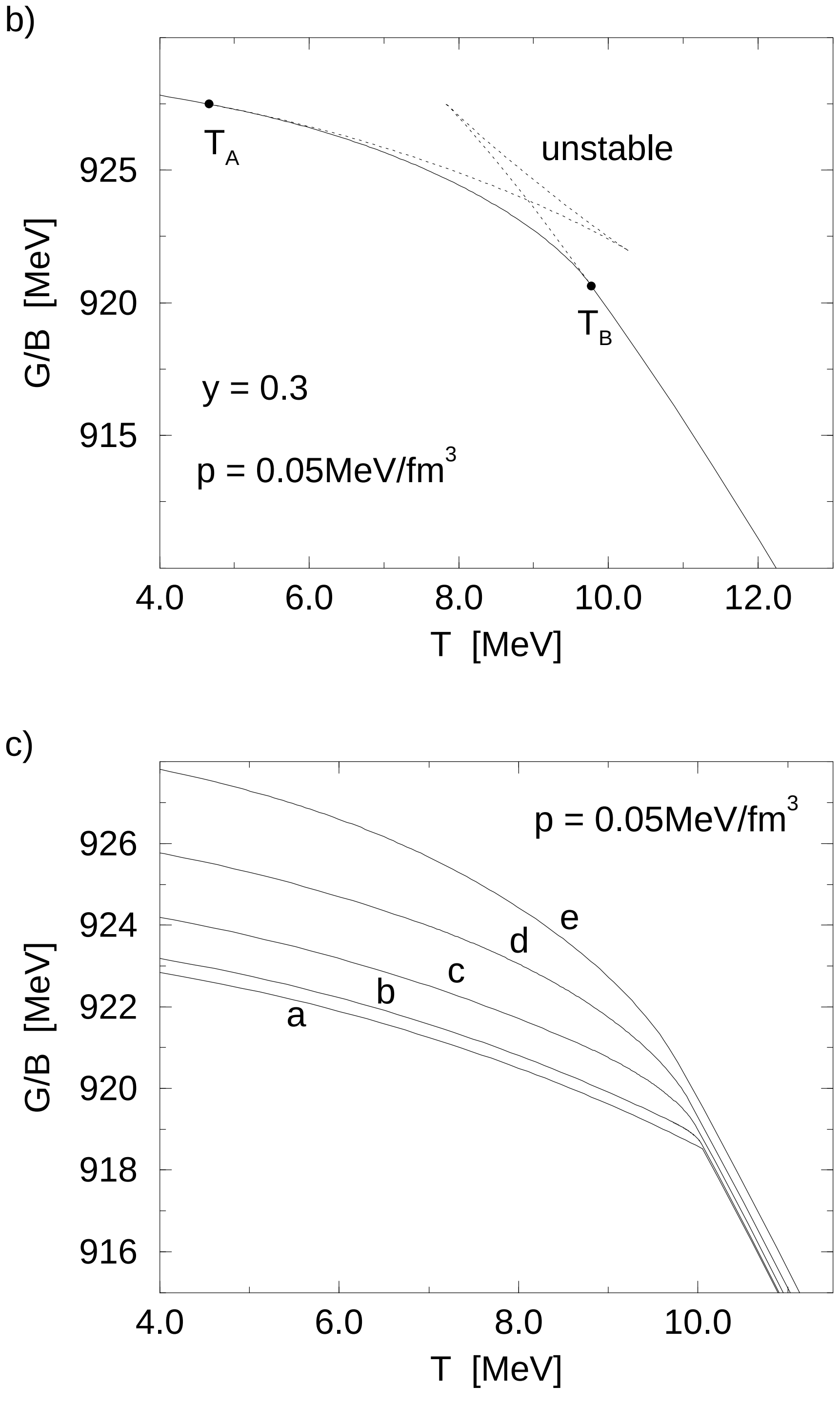

FIG. 14. Specific free enthalpy at constant pressure as a function of temperature. (a) Free enthalpy in symmetric matter. The solid curves represent the stable single-phase configurations, and their dotted extensions denote metastable systems. The matter is unstable along the remaining dotted curve. (b) Same as (a) for $y = 0.3$, showing that the free enthalpy is continuous throughout the phase transition. The points $T_A$ and $T_B$ denote the endpoints of the coexistence region. (c) Free enthalpy for various proton fractions. The values are $y = 0.5, 0.45, 0.4, 0.35, 0.3$ for curves $a$ through $e$, respectively.

In contrast, for matter with $N_p \neq N_n$ (Fig. 14b), the phase transition begins at a temperature $T_A$ and ends at $T_B > T_A$, leading to a completely smooth free enthalpy curve. The free enthalpy for various values of the proton fraction is shown in Fig. 14c to illustrate how the kink in $g$ develops as the system becomes more symmetric.

It is also of interest to examine the entropy per nucleon, which is defined by

$$s(T,p,y) = \frac{S}{B} = -\left(\frac{\partial g(T,p,y)}{\partial T}\right)_{p,y} . \tag{61}$$

The entropy in the transition region requires special attention, because only the *total* proton fraction is held constant in Eq. (61). Therefore

$$\begin{aligned} s(T,p,y) &= (1-\bar{\lambda})s(T,p,y') + \bar{\lambda}s(T,p,y'') + \left(\frac{\partial\bar{\lambda}}{\partial T}\right)_{p,y} [g(T,p,y') - g(T,p,y'')] \\ &\quad - (1-\bar{\lambda})\left(\frac{\partial y'}{\partial T}\right)_{p,y}\left(\frac{\partial g(T,p,y')}{\partial y'}\right)_{T,p} - \bar{\lambda}\left(\frac{\partial y''}{\partial T}\right)_{p,y}\left(\frac{\partial g(T,p,y'')}{\partial y''}\right)_{T,p} \\ &= (1-\bar{\lambda})s(T,p,y') + \bar{\lambda}s(T,p,y'') , \end{aligned} \tag{62}$$

where the last identity can be verified by observing that $y = (1-\bar{\lambda})y' + \bar{\lambda}y''$ and that the chemical potentials are identical in both phases. Note that in evaluating higher derivatives, *e.g.*, the heat capacity, similar care must be used to include the temperature dependence of $\bar{\lambda}$, $y'$, and $y''$.

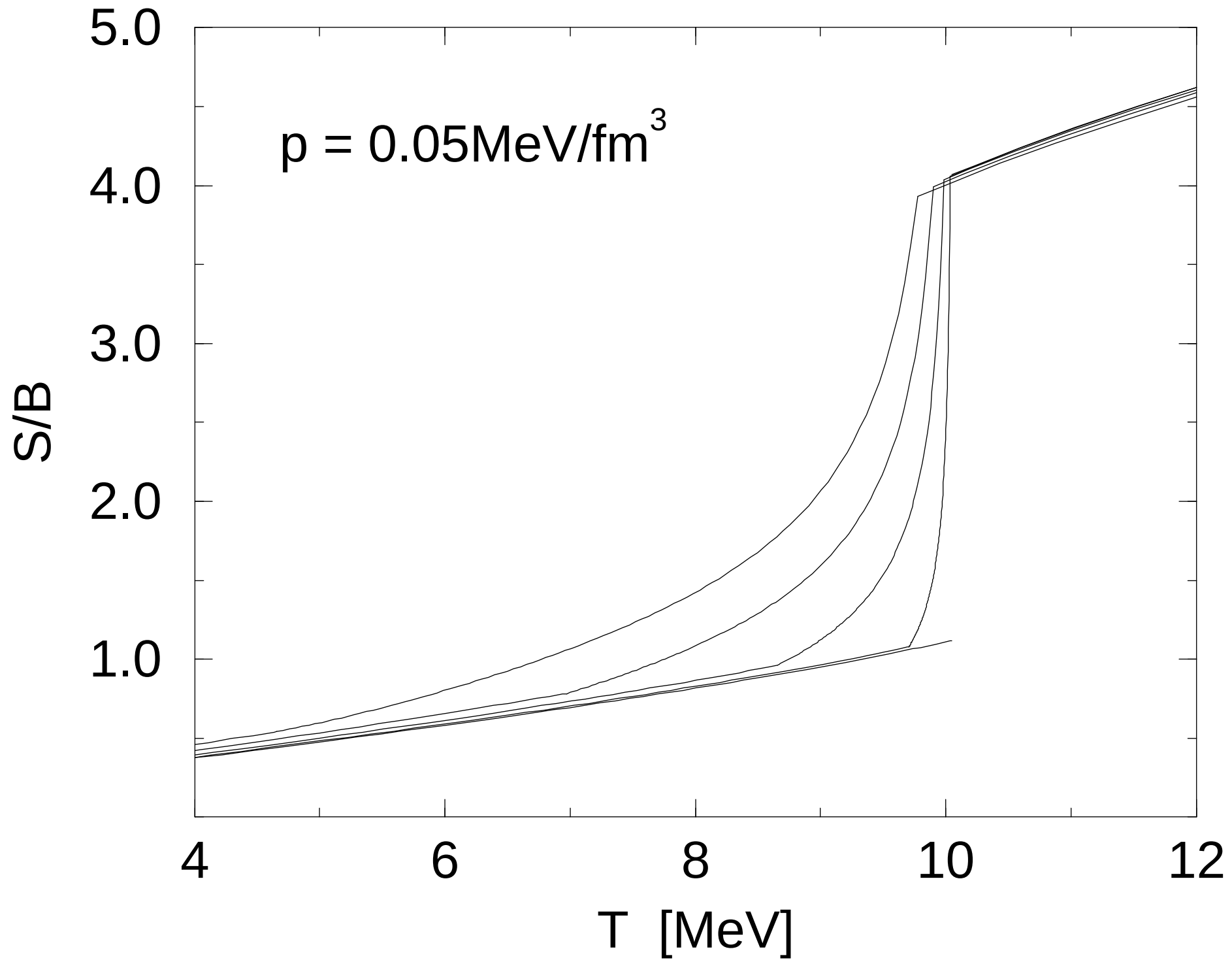

FIG. 15. Specific entropy as a function of temperature at constant pressure for various asymmetries. The curves have the same values of $y$ as in Fig. 14c, with $y = 0.5$ at the bottom.

Figure 15 shows the specific entropy for several different asymmetries. The kink in $g$ for the symmetric system gives rise to a discontinuity. In contrast, the entropy in an asymmetric system is *continuous*, but there are now kinks at the end points of the phase transition. By using

$$dQ = T ds \ , \tag{63}$$

we can calculate the amount of heat that has to be transferred to the system during the course of the transition. For the symmetric system, this yields the well-known expression for the latent heat:

$$Q_L = T(s_A - s_B) \ ,$$

which means that the all of the energy is used to convert the liquid into vapor.

The behavior of asymmetric matter is quite different. The temperature changes during the transition and so does the entropy, so that the integration becomes nontrivial:

$$Q = \int_{s_A}^{s_B} T(s) ds \ ,$$

and thus some fraction of the energy is used just to heat the system, as is evident from Fig. 15. The concept of latent heat is therefore not strictly applicable to the transition in the binary system. A well-known example of this behavior is the distillation of alcoholic beverages; the concentration of alcohol in the liquid changes throughout the distillation, which produces a change in the boiling point.

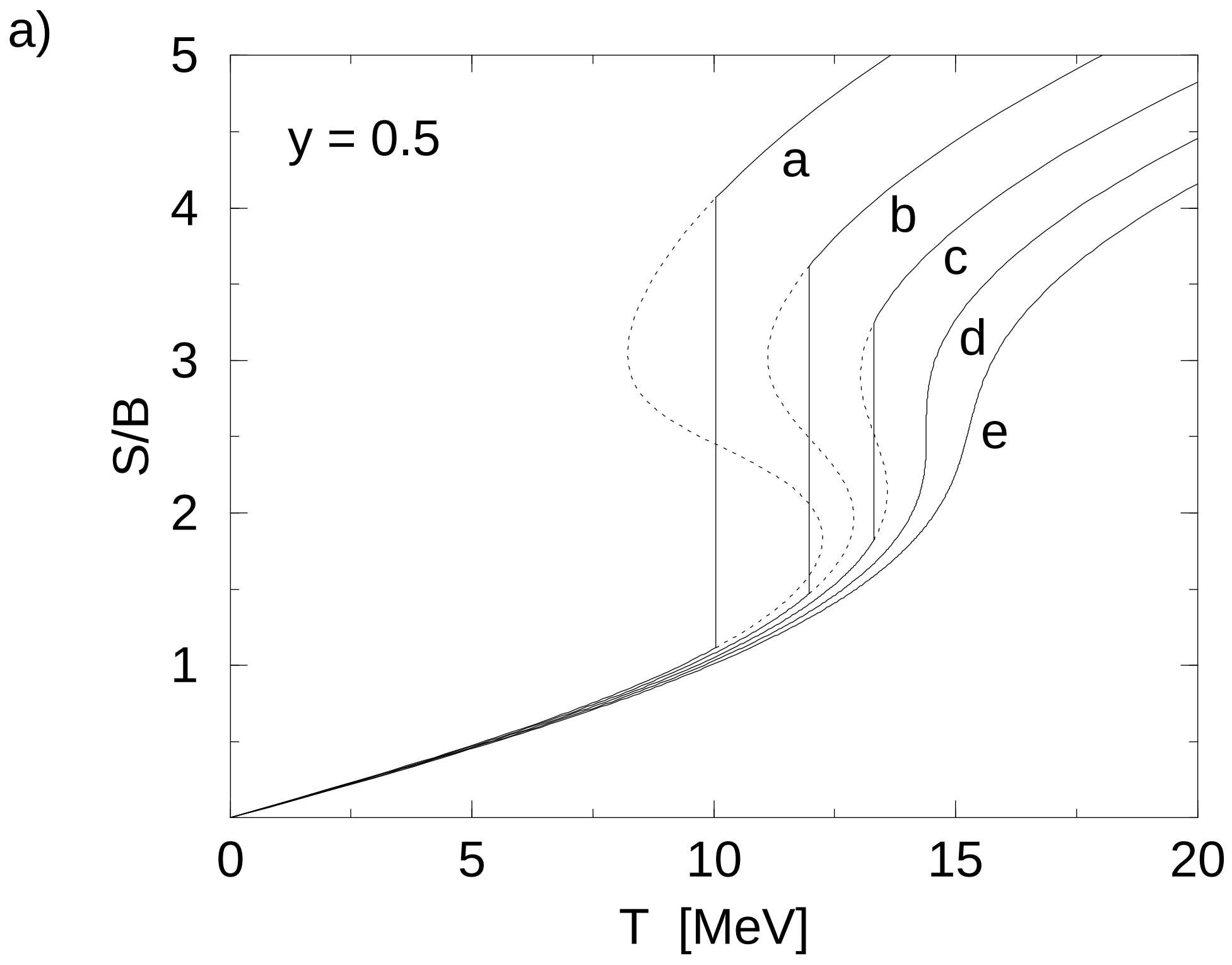

b)

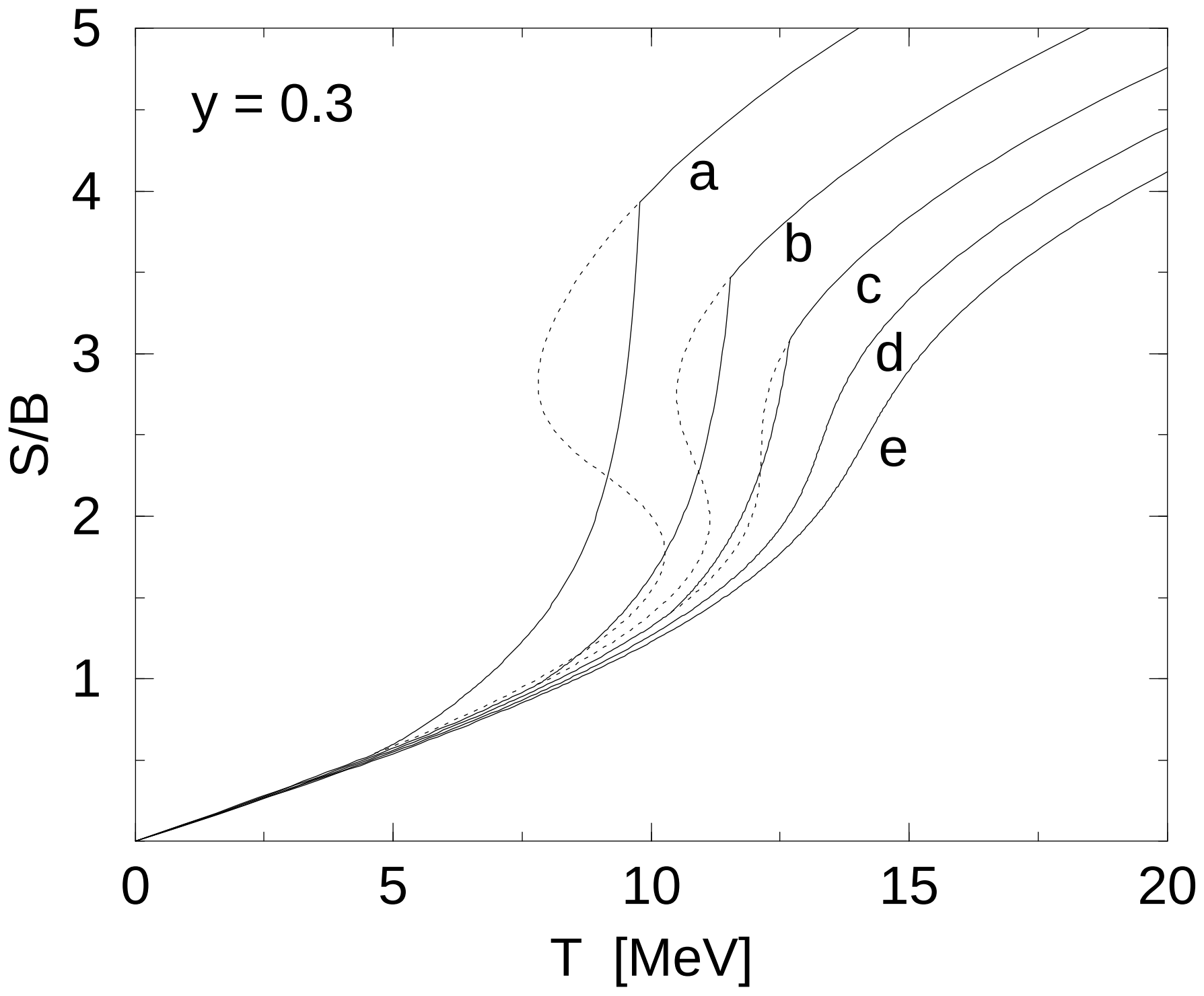

FIG. 16. Specific entropy as a function of temperature for different isobars. (a) Symmetric nuclear matter with $y = 0.5$. The pressures on the curves labeled $a$ through $e$ are $p = 0.05, 0.10, 0.15, 0.20, 0.25\,\mathrm{MeV/fm^3}$, respectively. Curve $d$ is at the critical pressure $p = p_c = 0.20\,\mathrm{MeV/fm^3}$. (b) Asymmetric nuclear matter with $y = 0.3$. The curves are labeled as in (a), and the critical pressure (curve $d$) is still $p_c = 0.20\,\mathrm{MeV/fm^3}$ (to two significant figures).

Figures 16a and 16b show the entropy as a function of temperature for several different isobars in symmetric and asymmetric matter, respectively. By examining the curves at the critical pressure $p_c$, one observes that $(\partial s/\partial T)_{p,y}$ becomes infinite in symmetric matter, while it remains finite and positive in the asymmetric case.

Finally, the distinct behavior of the binary system has an even more dramatic impact on the heat capacity per nucleon:

$$c_p = \frac{C_p}{N_p + N_n} = T \left( \frac{\partial s}{\partial T} \right)_{p,y} , \tag{64}$$

which is illustrated in Fig. 17. In symmetric matter, the discontinuity in the entropy produces an undefined heat capacity at the transition temperature. In contrast, in the general case, there are finite discontinuities in the heat capacity at the endpoints of the transition region, with finite values of $c_p$ in between. Thus, according to Ehrenfest's definition of phase transitions [46], the *first-order* phase transition in symmetric matter becomes a *second-order* transition in the asymmetric case. In particular, in the present model, the phase transition occurs over $\approx 1.5\,\mathrm{MeV}$ for matter that is 40% protons and over $\approx 5\,\mathrm{MeV}$ for matter that is 30% protons. The latter concentration might be obtained in energetic collisions with radioactive ion beams.

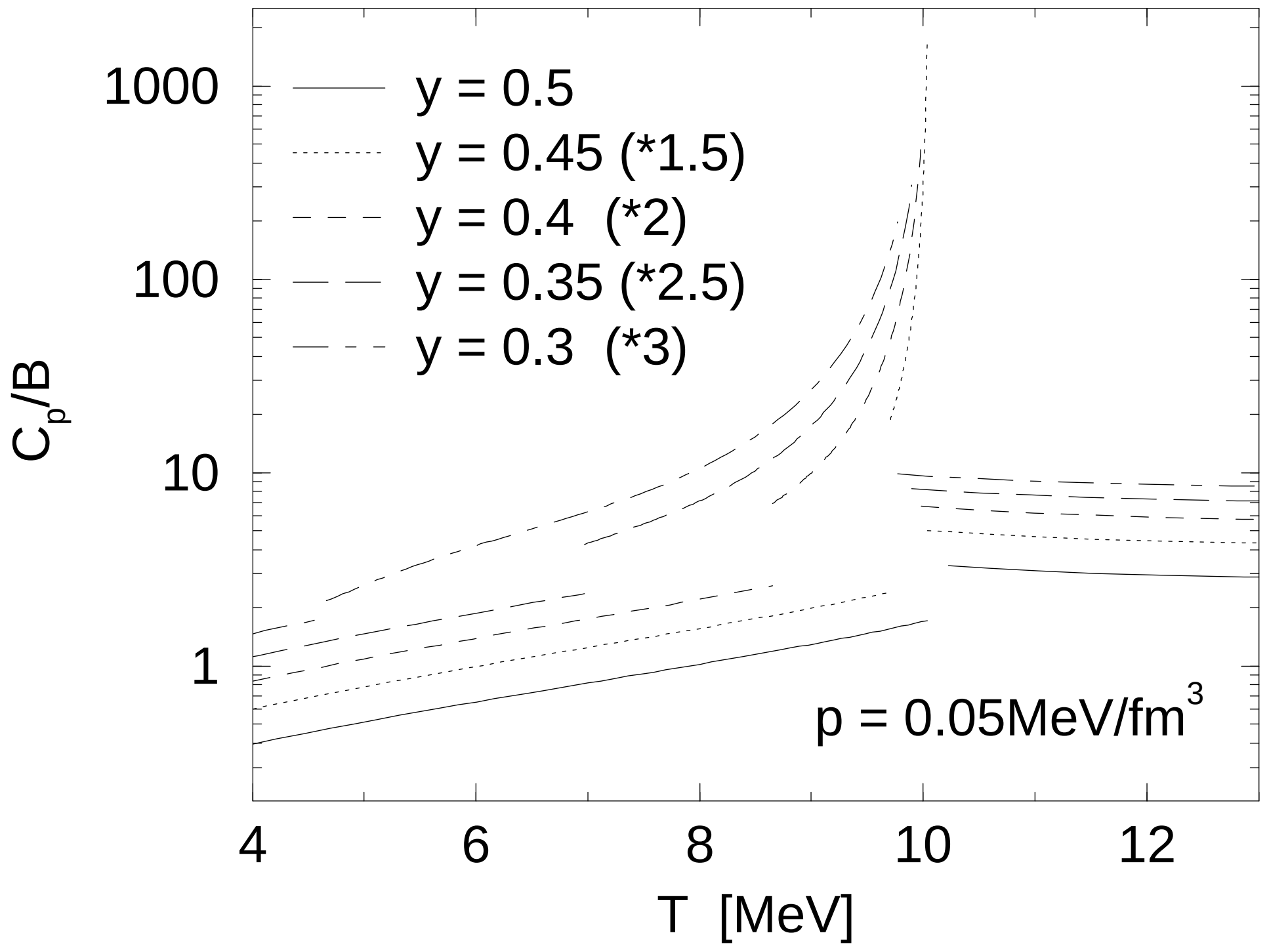

FIG. 17. Specific heat capacity as a function of temperature for several proton fractions. For clarity, the curves for asymmetric matter have been scaled by the factors indicated in parentheses. Note the logarithmic scale on the left.

## V. HEAVY-ION COLLISIONS

We consider the energetic collision of two heavy ions within our simple thermodynamic, hydrodynamic, and mean-field picture, and concentrate on the new features that arise as a function of proton fraction. Although our discussion of the theory will encompass the entire regime of $y < 0.5$, to obtain reasonable estimates for the empirical size of the new effects, we will restrict consideration to $0.3 \leq y \leq 0.5$, where the lower value might be obtainable with radioactive ion beams. We shall assume that the combined system is compressed and heated and ultimately reaches equilibrium at some finite temperature, density, and pressure. As is well known, the question of whether the system actually reaches such an equilibrium state is a difficult one, which we will not attempt to answer here. We simply assume that such a state arises and follow the subsequent expansion and cooling of the nuclear matter. The properties of the system can be deduced from Figs. 18 through 20, where the pressure is shown as a function of the baryon density for various proton fractions, temperatures, and specific entropies. Also indicated are the coexistence curves or binodals, which are determined by the Maxwell construction discussed earlier.

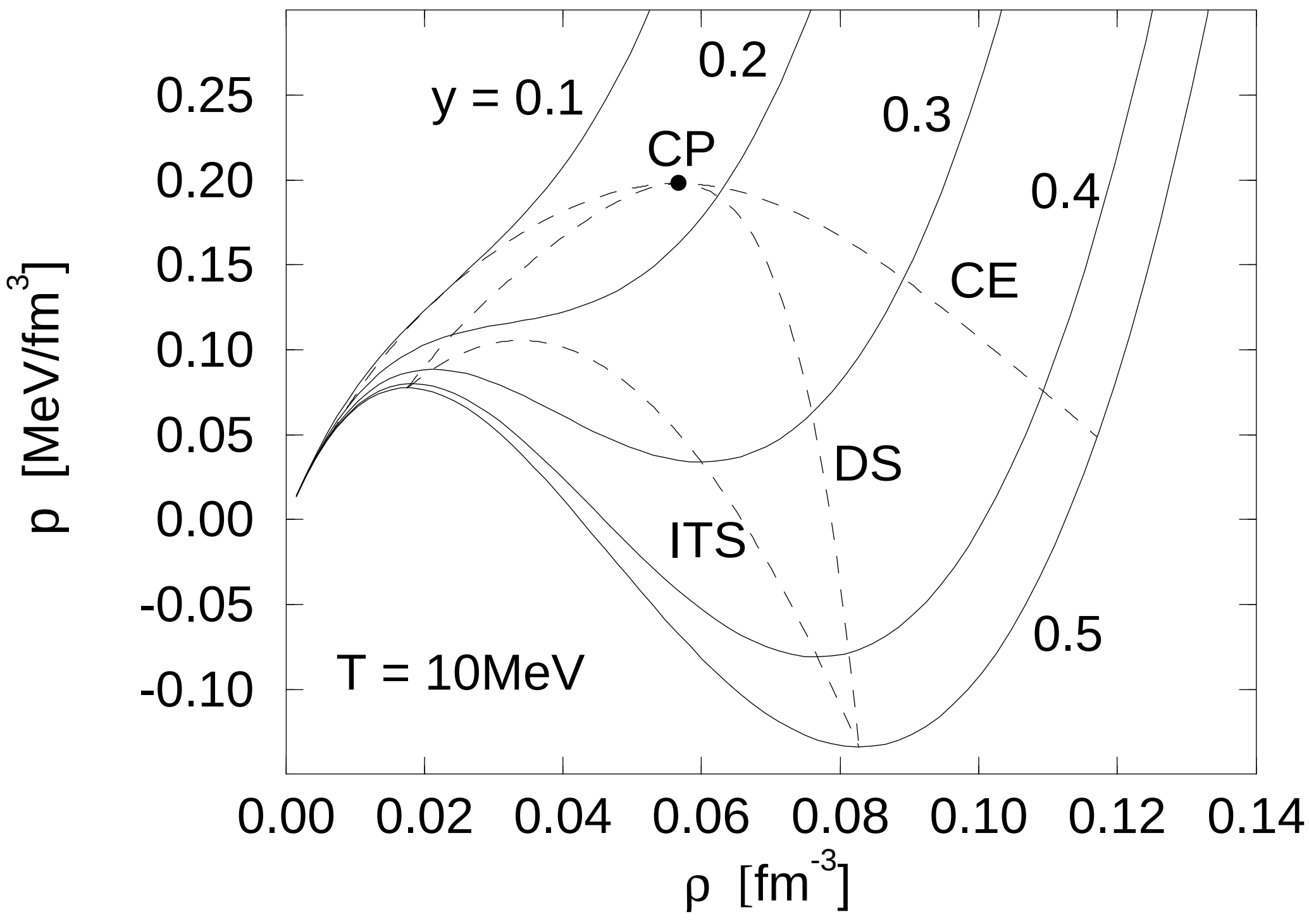

FIG. 18. Pressure as a function of density at fixed temperature for various $y$. The critical point (CP), coexistence curve (CE), diffusive spinodal (DS), and isothermal spinodal (ITS) are indicated. The dashed curves are discussed in the text.

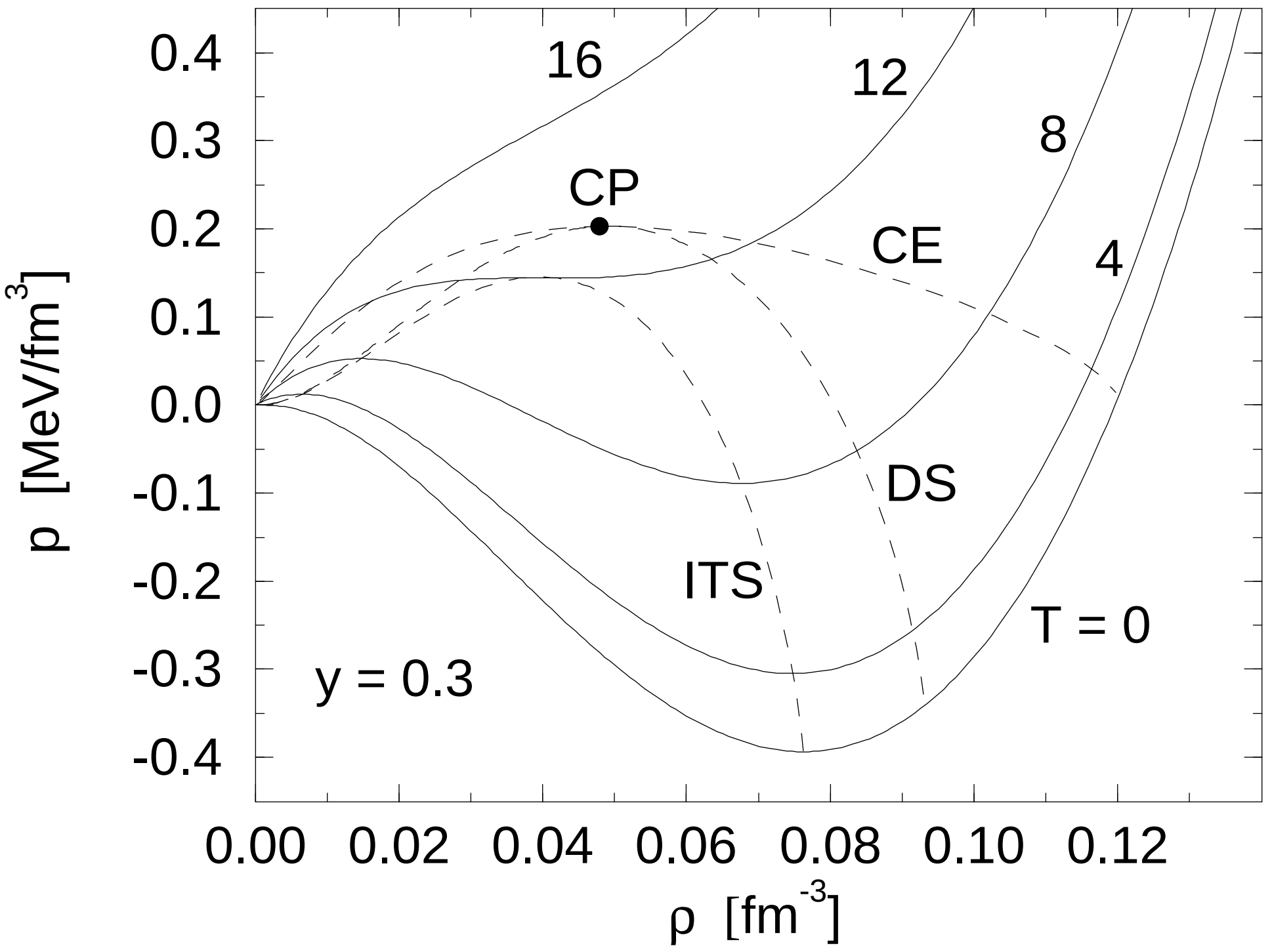

FIG. 19. Pressure as a function of density at fixed $y = 0.3$. The solid curves are labeled by the temperature (in MeV). The critical point (CP), coexistence curve (CE), diffusive spinodal (DS), and isothermal spinodal (ITS) are indicated. The dashed curves are discussed in the text.

The spinodals that determine the boundaries of the unstable region now generally come in three varieties, one arising from a mechanical instability, one arising from a thermal instability, and one arising from a diffusive (or chemical) instability. The conditions that determine the different spinodals will be discussed shortly. The region between the binodal and the most extensive spinodal contains metastable states, which correspond either to superheated liquid or supercooled (supersaturated) vapor.

The metastable states play a key role in understanding the phase transition. In stable configurations, Eq. (8) is satisfied, so that

$$\Delta\mathcal{F}_b \equiv (1-\lambda)\mathcal{F}(T,\rho',y') + \lambda\mathcal{F}(T,\rho'',y'') - \mathcal{F}(T,\rho,y) > 0 \tag{65}$$

holds for all densities that obey the conservation laws

$$\begin{aligned} \rho &= (1-\lambda)\rho' + \lambda\rho'' \ , \\ y\rho &= (1-\lambda)y'\rho' + \lambda y''\rho'' \ . \end{aligned} \tag{66}$$

In the metastable region, two phases can be found that produce $\Delta\mathcal{F}_b < 0$, but the energy barrier $\Delta\mathcal{F}_b$ remains positive for small variations in $\rho$ and $y$ about the single-phase values. Thus the free-energy density is *locally* convex, and the energy barrier prevents phase separation due to infinitesimal fluctuations. Eventually, $\Delta\mathcal{F}_b$ becomes negative as the system enters the labile region, signaling the instability even to infinitesimal fluctuations. We postpone a discussion of the various mechanisms that can produce the phase separation and concentrate here on the behavior of the free energy in the region of the spinodals.

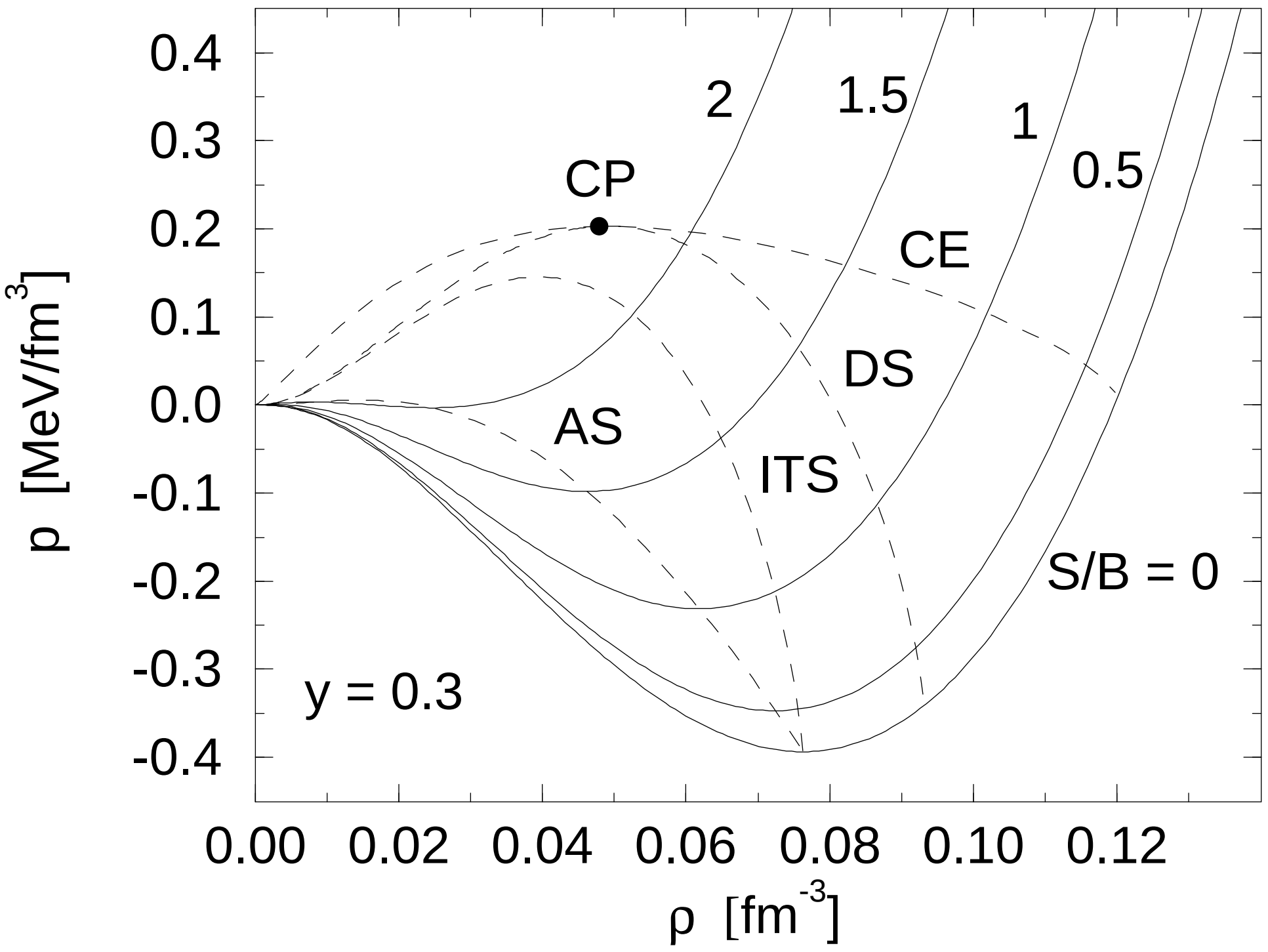

FIG. 20. Pressure as a function of density at fixed $y = 0.3$. The solid curves are labeled by the specific entropy. The critical point (CP), coexistence curve (CE), diffusive spinodal (DS), isothermal spinodal (ITS), and adiabatic spinodal (AS) are indicated. The dashed curves are discussed in the text.

According to the discussion in Sec. II, we have

$$\Delta\mathcal{F}_b > 0 \quad \text{for} \qquad \left(\frac{\partial p}{\partial\rho}\right)_{T,y} > 0 \quad \text{and} \quad \left(\frac{\partial\mu_p}{\partial y}\right)_{T,p} > 0 \tag{67}$$

in the stable and metastable region. The content of these relations becomes more transparent if we consider small fluctuations around the equilibrium density and concentration:

$$\begin{aligned} \rho' &= \rho + \Delta\rho' \ , \quad y' = y + \Delta y' \ , \\ \rho'' &= \rho + \Delta\rho'' \ , \quad y'' = y + \Delta y'' \ . \end{aligned} \tag{68}$$

An expansion through second order in small quantities produces

$$\Delta\mathcal{F}_b \approx \frac{1-\lambda}{2\lambda}\left\{\frac{1}{\rho}\left(\frac{\partial p}{\partial\rho}\right)_{T,y}(\Delta\rho')^2 + \rho\left[\left(\frac{\partial\mu_p}{\partial y}\right)_{T,\rho} - \left(\frac{\partial\mu_n}{\partial y}\right)_{T,\rho}\right](\Delta y')^2 \right.$$
$$\left. +2\rho\left[\left(\frac{\partial\mu_p}{\partial\rho}\right)_{T,y} - \left(\frac{\partial\mu_n}{\partial\rho}\right)_{T,y}\right]\Delta y'\Delta\rho'\right\} + \dots \ , \tag{69}$$

where $\Delta\rho''$ ($\Delta y''$) has been eliminated in favor of $\lambda$ and $\Delta\rho'$ ($\Delta y'$) using Eq. (66). Consistent with our general discussion, this bilinear form will be positive if

$$\left(\frac{\partial p}{\partial \rho}\right)_{T,y} > 0 \tag{70}$$

and if

$$\left(\frac{\partial p}{\partial \rho}\right)_{T,y}\left[\left(\frac{\partial \mu_p}{\partial y}\right)_{T,\rho} - \left(\frac{\partial \mu_n}{\partial y}\right)_{T,\rho}\right] - \rho^2\left[\left(\frac{\partial \mu_p}{\partial \rho}\right)_{T,y} - \left(\frac{\partial \mu_n}{\partial \rho}\right)_{T,y}\right]^2$$
$$= \frac{1}{1-y}\left(\frac{\partial \mu_p}{\partial y}\right)_{T,p}\left(\frac{\partial p}{\partial \rho}\right)_{T,y} > 0 \ . \tag{71}$$

Therefore, as long as the conditions (70) and (71) are valid, the system remains on a parabolic free-energy surface, with a minimum at $\Delta\rho = \Delta y = 0$. The spinodals are defined by the densities at which these inequalities become invalid. As is clear from Fig. 18, there are generally *four* such densities of interest in an isothermal expansion, corresponding to the intersections of the isotherm with the spinodals. Each of these four densities marks a change in the sign of the derivative $(\partial\mu_p/\partial y)_{T,p}$. At the largest density, this derivative changes sign from positive to negative, indicating the onset of the diffusive instability; nevertheless, the compressibility remains positive, so that only the second inequality (71) is violated. This also implies that the free-energy density now has a saddle point, being stable against density fluctuations at fixed $y$, but unstable against fluctuations in concentration at fixed pressure. Through continued expansion, one next encounters the mechanical instability, at which point $(\partial p/\partial\rho)_{T,y}$ becomes negative, but $(\partial\mu_p/\partial y)_{T,p}$ now becomes positive again, so that *both* conditions (70) and (71) are violated. Thus the free-energy surface still has a saddle point, but it has rotated in the $(\rho, y)$ plane. As the density decreases further, one eventually reaches densities where the derivatives change sign in reverse order, producing the remaining two intersections of the spinodals with the isotherm.

The conclusion from this analysis is that the spinodal structure is qualitatively different in the asymmetric system than in the symmetric one. There are now *two* types of fluctuations: one corresponding to changes in $\rho$ (isoscalar) and one corresponding to changes in $y$ (isovector). Due to the form of the nuclear symmetry energy, the diffusive spinodal (DS) encloses more of the configuration space than does the isothermal spinodal (ITS). The DS passes through the critical point and includes all densities where there is also a mechanical instability, which occur in the region bounded by the ITS. *Therefore the diffusive instability defines the relevant spinodal for the asymmetric system.* The mechanical instabilities are restricted to isotherms with $y > y_*$, where $y_*$ is defined by the inflection point

$$\left(\frac{\partial p}{\partial \rho}\right)_{T,y_*} = \left(\frac{\partial^2 p}{\partial \rho^2}\right)_{T,y_*} = 0 \ .$$

As expected, both spinodals coincide at $y = 0.5$, where only mechanical instability is possible. Note also that there are isotherms that pass through the metastable region and never intersect either spinodal. These isotherms, along which retrograde condensation is possible, never become labile and allow for the system to evolve completely through the metastable region, if the process is carried out carefully enough (or fast enough).

So far, we have considered only isothermal processes, but in fact, the equilibrium evolution of the system will depend strongly on the variables that actually remain fixed during the process. As we have noted, it is not clear that *any* thermodynamic variables can be assigned to realistic situations encountered in warm, expanding nuclei. Nevertheless, intranuclear cascade calculations [8,14] suggest that this expansion will be isentropic. To discuss isentropic (adiabatic) processes, we must generalize our formalism slightly. The appropriate state function to describe trajectories at constant specific entropy $s = S/B$ is the energy/baryon, $E/B = e(s, v, y)$, where $v = 1/\rho$. The corresponding stability condition can be formulated as:

$$e(s,v,y) < (1-\bar{\lambda})e(s',v',y') + \bar{\lambda}e(s'',v'',y'') \ , \tag{72}$$

with

$$s = (1-\bar{\lambda})s' + \bar{\lambda}s'' \ , \quad v = (1-\bar{\lambda})v' + \bar{\lambda}v'' \ , \quad y = (1-\bar{\lambda})y' + \bar{\lambda}y'' \ , \tag{73}$$

where $\bar{\lambda}$ [see Eq. (60)] determines the fraction of particles in each phase.

As in the case of the free-energy density, the global criterion Eq. (72) can be reformulated in terms of local conditions, namely,

$$\left(\frac{\partial p}{\partial \rho}\right)_{s,y} > 0 \ , \tag{74}$$

$$\frac{1}{c_p}\left(\frac{\partial p}{\partial \rho}\right)_{s,y} > 0 \ , \tag{75}$$

$$\frac{1}{c_p}\left(\frac{\partial p}{\partial \rho}\right)_{s,y}\left(\frac{\partial \mu_p}{\partial y}\right)_{T,p} > 0 \ , \tag{76}$$

where $c_p$ is defined in Eq. (64). The first two inequalities are relevant for both symmetric (one-component) and asymmetric (two-component) systems, and the third embodies the diffusive or chemical stability criterion that arises for asymmetric matter. Thus there will generally be three spinodals, as indicated in Fig. 20.

To connect this to our earlier discussion of isothermal processes, we use the relation [47]

$$\frac{1}{c_p}\left(\frac{\partial p}{\partial \rho}\right)_{s,y} = \frac{1}{c_v}\left(\frac{\partial p}{\partial \rho}\right)_{T,y} \tag{77}$$

to rewrite Eqs. (75) and (76) as

$$\frac{1}{c_v}\left(\frac{\partial p}{\partial \rho}\right)_{T,y} > 0 \ , \tag{78}$$

$$\frac{1}{c_v}\left(\frac{\partial p}{\partial \rho}\right)_{T,y}\left(\frac{\partial \mu_p}{\partial y}\right)_{T,p} > 0 \ . \tag{79}$$

The equivalence of these two ways of writing the stability criteria implies that the stability boundaries are *independent* of the actual process. Moreover, since in our model (see also Ref. [19]), $c_v = T(\partial s/\partial T)_{v,y}$ is always positive, the conditions (78) and (79) reduce to those discussed earlier in Eqs. (70) and (71). Thus, phase separations in isothermal processes in symmetric matter occur only due to mechanical instabilities (density fluctuations), as determined by the ITS. In asymmetric systems, the final stability condition is also relevant, leading to diffusive instabilities and the appearance of the DS as well.

In contrast, $c_p$ undergoes several sign changes in an adiabatic expansion, so that all three criteria (74)–(76) must be considered. One sign change accompanies a sign change in the isothermal compressibility [or $(\partial p/\partial \rho)_{T,y}$], and one accompanies a corresponding change in the adiabatic compressibility [or $(\partial p/\partial \rho)_{s,y}$]. [4] The resulting situation is depicted in Fig. 20. The outermost spinodal (DS) defines where Eq. (76) is violated, which occurs due to the diffusive instability, since both $c_p$ and the compressibilities are all positive here. The middle spinodal is determined by condition (75), which is violated when $c_p$ becomes negative; because of Eq. (77), this curve is *identical* to the ITS, since $c_v$ and $(\partial p/\partial \rho)_{s,y}$ are still positive, so $(\partial p/\partial \rho)_{T,y}$ must also become negative. In other words, the ITS encloses the region of mechanical instability in an isothermal expansion and *also* encloses the region of thermal instability in an adiabatic expansion. The mechanism for this thermal instability is interesting, because $c_p$ changes sign by passing through infinity rather than zero. Thus a small thermal fluctuation *at constant pressure* near the ITS creates a density fluctuation, but leaves the temperature unchanged (since $c_p$ is very large); since the isothermal compressibility becomes negative, the system is unstable to the induced density fluctuation.

Finally, the innermost curve is the adiabatic spinodal (AS), which determines when Eq. (74) is violated, signaling the onset of mechanical instability in the adiabatic process. Here $c_p$ becomes positive again by Eq. (77), since both compressibilities are negative here. Note also that, as discussed earlier, $(\partial \mu_p/\partial y)_{T,p}$ changes sign from positive to negative at the DS and then changes back again at the ITS, so that once any stability criterion (74)–(76) is violated, it remains invalid as one proceeds deeper into the unstable region.

Just as we found in the case of isothermal processes, the DS separates the metastable and labile regions in isentropic processes. Configurations that indicate sign changes of the compressibilities and of $c_p$ are included by this boundary. We also observe that even in symmetric nuclear matter, the ITS determines the boundary of the labile region in *both* isothermal and isentropic processes, which has often been overlooked in previous studies [7,8,11,12,36,48]. Although this result apparently implies that the AS is irrelevant, this conclusion is premature, since the actual mechanism for spinodal decomposition depends on the relative rate of thermal and mechanical fluctuations as the warm matter expands through the spinodal region.

Finally, because of the numerous relations between the thermodynamic variables and their derivatives enumerated above, correct results will be obtained only if the underlying calculation of the nuclear equation of state is thermodynamically consistent. This is known to be true for the relativistic MFT used here [22,36], but consistency is difficult to maintain in more sophisticated approximations to the relativistic many-body problem [49].

Armed with this understanding of the spinodal structure, we turn now to the evolution of the warm matter.

---

[4]Note that the usual definition of the isothermal compressibility [47], $\kappa_T \equiv -(\partial V/\partial p)_T/V$ can be rewritten as $\rho(\partial p/\partial \rho)_T = 1/\kappa_T$. Analogous expressions hold at fixed entropy.

a)

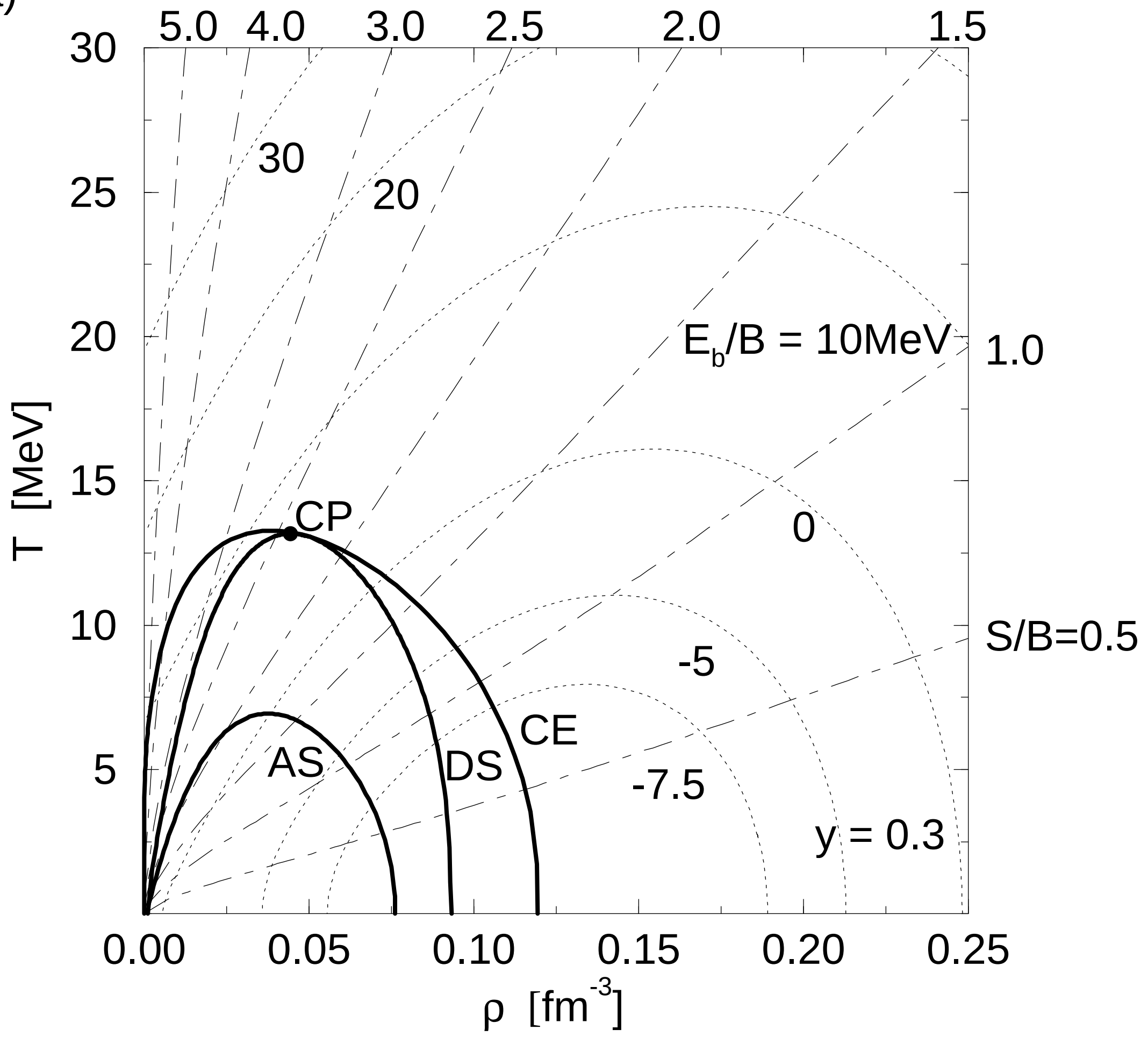

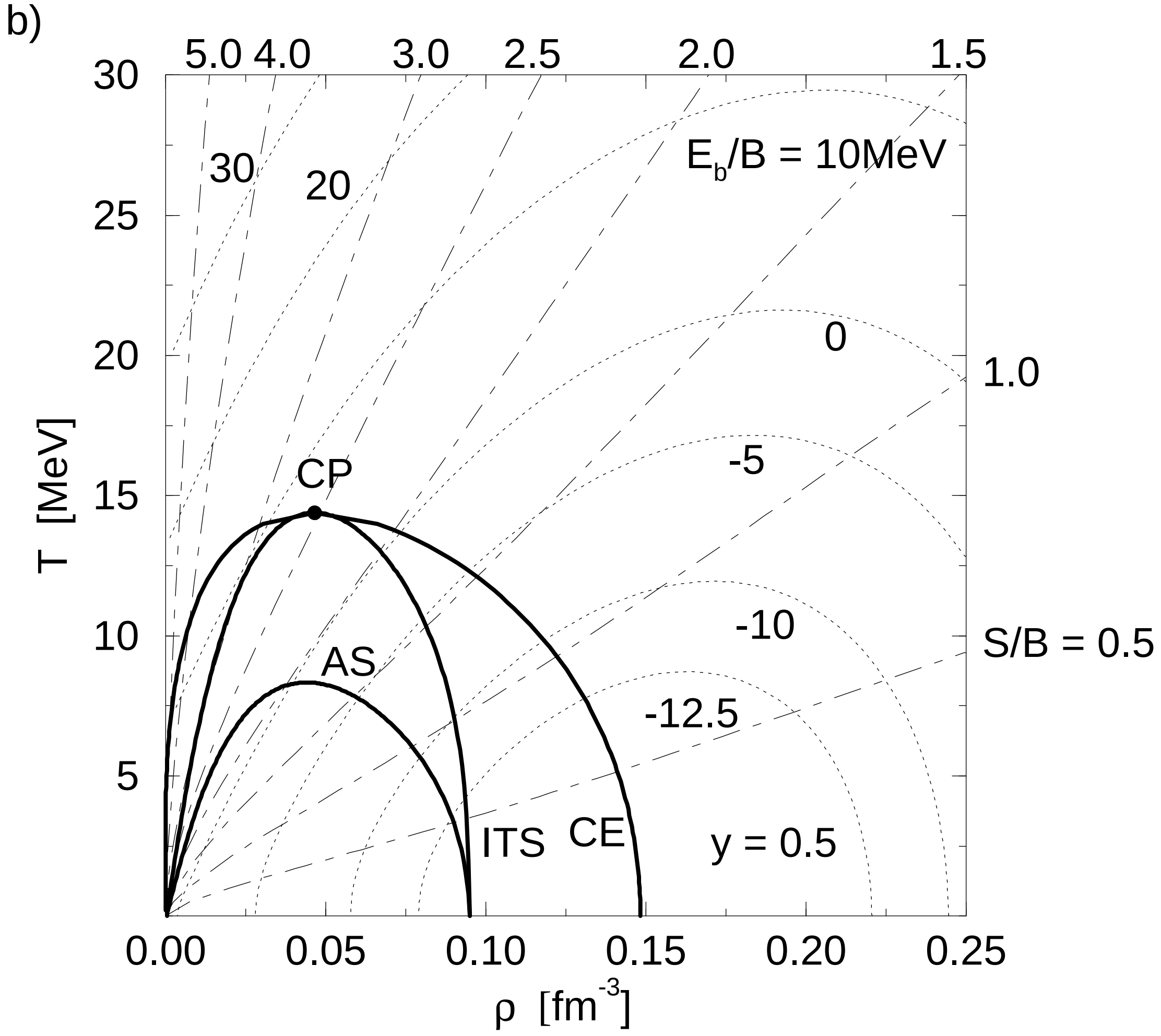

FIG. 21. Properties of nuclear matter as functions of temperature and density. The dotted curves are contours of equal energy/baryon (in MeV), and adiabats are shown as dot-dashed curves. The solid lines denote the coexistence curve (CE), the diffusive spinodal (DS), the isothermal spinodal (ITS), and the adiabatic spinodal (AS). Part (a) shows results for asymmetric matter with $y = 0.3$ and part (b), for symmetric matter ($y = 0.5$).

The relation between the evolution and the thermodynamic variables can be studied using Fig. 21, which shows the properties of the matter for this equation of state in the $T, \rho$ plane for two different values of $y$. Equipotential surfaces and adiabats are also indicated. The relations used to determine the nuclear matter properties at very low densities are given in the Appendix.

The solid curves in the lower-left corners of Figs. 21a,b determine the phase behavior. The outer curve labeled CE is the coexistence curve, which is simply a section through the binodal surface at fixed $y = 0.3, 0.5$. Inside this curve, the stable configuration is a mixture of liquid and gas. Also shown is the diffusive spinodal (DS), which is determined by the condition $(\partial\mu_p/\partial y)_{T,p} = 0$, as discussed earlier. Between the spinodal and the CE, the system can exist in metastable superheated or supersaturated states.

There are several qualitative differences between the curves for asymmetric matter and for symmetric matter (see also Fig. 8 in Ref. [36]). First, the most extensive spinodal is defined by the diffusive instability rather than by thermal or mechanical instabilities. (We note that mechanical instability is often assumed to be the relevant one even in asymmetric systems.) Second, as is evident from Fig. 8, the critical temperature $T_c(y)$ is not unique, but varies with the concentration along the LCP. Finally, the maximum temperature of phase separation does not occur at the critical point, but rather along the LMA; it is apparent from Fig. 21a that this occurs at a lower density than the density at the critical point. We observe, however, that although these differences exist in principle, for physically accessible systems, the temperature differences are small. For example, in the present model, the critical temperature changes from $T_c(0.5) = 14.4\,\mathrm{MeV}$ to $T_c(0.3) = 13.1\,\mathrm{MeV}$, and $T_{\max}(0.3) - T_c(0.3) \approx 0.2\,\mathrm{MeV}$. Since statistical fluctuations near the critical point are expected to prevent the determination of the critical temperature with an uncertainty smaller than 1 or 2 MeV [16], it is unlikely that these small temperature differences can be directly observed. Nevertheless, a trend in the observable signals characteristic of the critical point as a function of increasing asymmetry may be detectable.

The evolution of the system is a complicated process that involves hydrodynamic flow and expansion, together with dissipative effects and nonequilibrium processes like nucleation and fragmentation. As the system expands, internal energy is transformed into collective motion, which manifests itself as local flow velocity. If the expansion is highly damped, the energy of motion is rapidly transformed back into internal energy; the resulting expansion is therefore slow and proceeds along an equipotential surface. The expansion continues until $p = 0$, after which the warm nuclear fluid remains at rest and evaporates particles until it is cool. In contrast, if the motion is undamped, the expansion carries the system past hydrostatic equilibrium, where it begins to slow down as the energy of motion is returned to internal energy. Since there is no dissipation, the expansion is isentropic, and the motion is bounded by the equipotential surface of the initial hot configuration.

As noted earlier, intranuclear cascade calculations [8,14] suggest that the expansion will be isentropic. The system will expand adiabatically into the coexistence region and become either superheated (for small specific entropy) or supersaturated (for large specific entropy). If the expansion is slow enough to allow for nucleation, bubbles of gas form or the system separates into droplets and vapor. If the nucleation is relatively slow, which is more likely, and the expansion halts before a spinodal is reached, the direction of motion is reversed and the system vibrates, ultimately evaporating neutrons to cool down. In contrast, if the system crosses a spinodal, fragmentation occurs; the relevant mechanism for the decomposition (that is, which spinodal is the relevant one) depends on the relative rates for isovector, isoscalar, and thermal fluctuations.

With our simple picture that focuses on equilibrium states, it is impossible to say anything definitive about the dynamics of the phase separation. This is especially true since it is unlikely to occur through a sequence of equilibrium configurations and will involve instead superheating, supercooling, or spinodal decomposition. Nevertheless, the dynamics of the phase separation is intimately linked to the character of the phase transition; for example, the critical exponents are expected to play an important role [6,16]. Since the phase transition in asymmetric matter is not of the usual van der Waals type, there could be observable consequences in the signals used to detect the phase separation, such as the distribution of emitted mass fragments. In particular, unlike the van der Waals case, the pressure, temperature, density, and concentration of the gas and liquid phases can change throughout the phase separation process. Moreover, since the spinodal decomposition occurs generally through a diffusive instability, any microscopic model of the fragmentation process must allow the gas and liquid phases to have different concentrations. The extension of existing models of nucleation, fragmentation, or percolation (see, for example, Refs. [16,50–52]) to allow for this new degree of freedom is an important topic for future study.

Finally, we discuss how the properties of the phase transition depend on reasonable changes in the nuclear matter symmetry energy and compressibility.

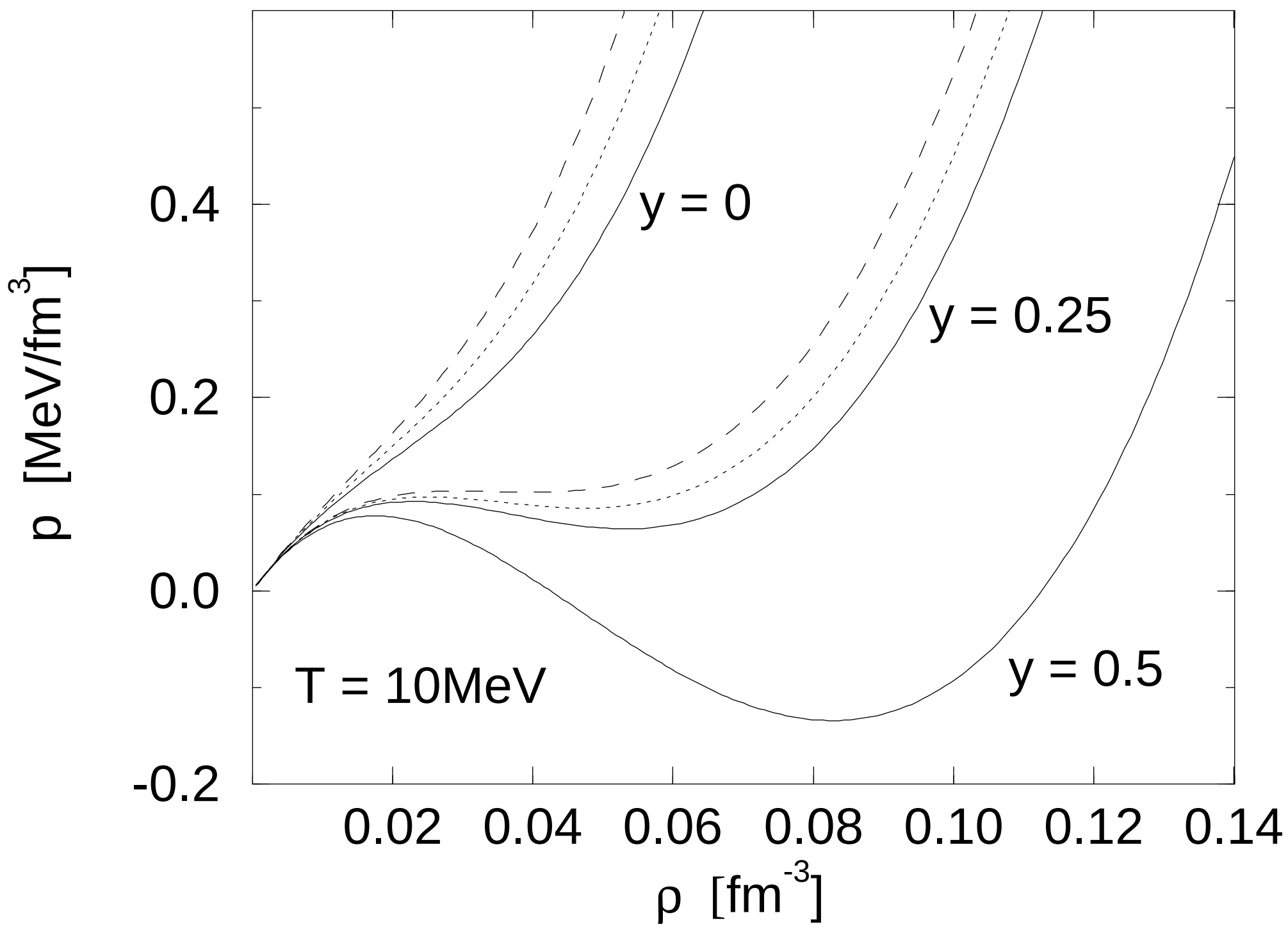

FIG. 22. Pressure as a function of density at fixed temperature for different values of the proton fraction. The nuclear symmetry energy $a_4 = 30, 35, 40\,\mathrm{MeV}$ for the solid, dotted, and dashed curves, respectively.

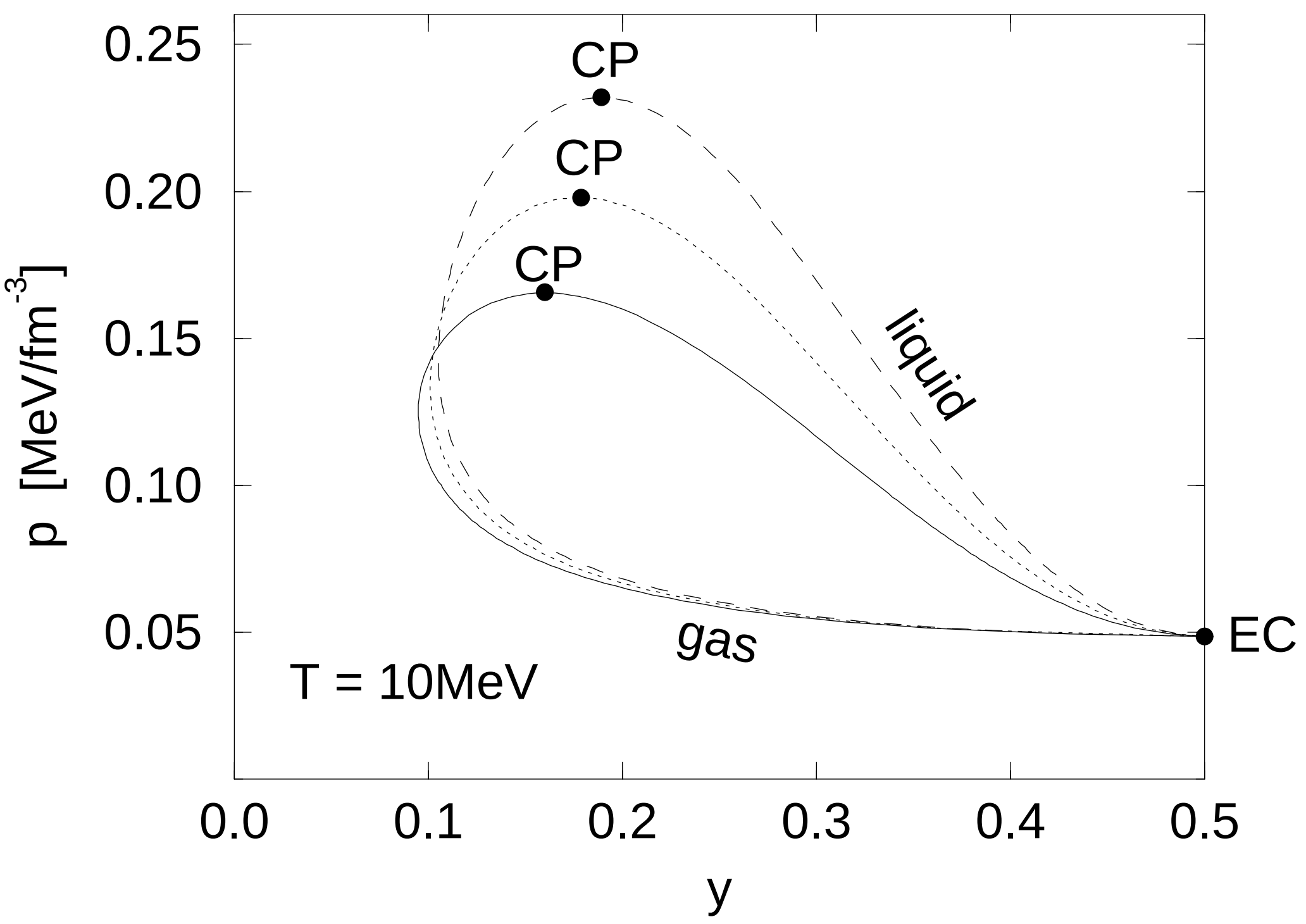

FIG. 23. Binodal sections at $T = 10\,\text{MeV}$ for different $a_4$. The nuclear symmetry energy $a_4 = 30, 35, 40\,\text{MeV}$ for the solid, dotted, and dashed curves, respectively, and the critical points (CP) are indicated in each case.

The symmetry energy can be varied by changing the coupling $C_\rho^2$, and we chose the values $a_4 =$ 30, 35, and 40 MeV for illustration. (All other couplings in Table I were held fixed.) In Figs. 22 and 23, we show the variations in the pressure and in the binodal surface as the symmetry energy is changed. As expected, as the symmetry energy increases, the minimum $y$ (maximal asymmetry) at which phase separation occurs increases, but only by roughly 10%. The critical pressure at fixed temperature increases somewhat more ($\approx 40\%$), but the shape of the binodal surface is qualitatively similar; the increased symmetry energy evidently has more effect on the gas phase than on the liquid phase. This result is consistent with our earlier discussion that it is energetically favorable for nuclear matter to separate into a less asymmetric liquid and a more asymmetric gas. As the symmetry energy increases, one finds a larger region of configuration space where this phase separation is favorable, particularly for retrograde condensation. The critical temperature at $y = 0.35$ decreases from $T_c(0.35) = 13.77$ MeV for $a_4 = 30\,\text{MeV}$, to $T_c(0.35) = 13.71$ MeV for $a_4 = 35\,\text{MeV}$, to $T_c(0.35) = 13.66$ MeV for $a_4 = 40\,\text{MeV}$, which is a relatively small amount. These modest changes as the symmetry energy is varied support the claim made in the Introduction that our simple model of the isovector mean-field dynamics is adequate for a discussion of nucleus-nucleus collisions.

We also studied the binodal surface for compressibilities $K_V^{-1} =$ 200, 250, and 300 MeV. (We return to our original value of the symmetry energy, $a_4 = 35\,\text{MeV}$.) The qualitative structure of the surface was unchanged, and the most significant feature was a shift in the critical temperature of symmetric matter from $T_c(0.5) = 13.35\,\text{MeV}$ at $K_V^{-1} = 200\,\text{MeV}$ to $T_c(0.5) = 15.75\,\text{MeV}$ at $K_V^{-1} = 300\,\text{MeV}$. Changes in $T_c$ for other values of $0.3 \leq$ $y \leq 0.5$ were similar. As noted earlier, these variations of 1 or 2 MeV are unlikely to be observable, due to statistical fluctuations near the critical point. We therefore conclude that the observation of the liquid–gas phase transition in nuclear collisions is unlikely to provide definitive information on the nuclear matter symmetry energy or compressibility, although the critical temperature is apparently more sensitive to the latter.

## VI. SUMMARY

In this paper we studied the liquid–gas phase transition in warm, low-density nuclear matter as a function of the proton fraction. The new ingredient in the analysis was the careful treatment of the two conserved charges, baryon number and isospin, which shows that the phase transition does not exhibit the usual van der Waals behavior. An examination of the stability criteria on the free energy, together with the charge conservation laws and Gibbs' criteria for phase equilibrium, reveals that the system can be specified by the same number of input variables (temperature and densities) regardless of the number of phases. The equilibrium conditions determine the region in parameter space where separation into two phases is energetically favorable, as well as the boundary of this region, the binodal surface. This surface is two-dimensional for a two-phase system with two conserved charges, in contrast to the familiar one-dimensional surface when there is but one conserved charge, and this leads to qualitatively new behavior, such as retrograde condensation. For specified input variables, the number of equations determining the properties of the phases is equal to the number of unknowns, allowing for an unambiguous Maxwell construction of the equilibrium state at any point during the transition. In contrast to the usual van der Waals case, we learned that in general, the pressure, temperature, density, and concentration of both the gas and liquid phase can vary throughout the transition. Moreover, both the Gibbs free energy and entropy are continuous throughout the transition, showing that it is *second-order* (by Ehrenfest's definition) rather than first-order.

To apply these results to nuclear matter, we used a relativistic mean-field model involving the interaction of baryons with scalar and vector fields. This model allows for an accurate description of the bulk properties of nuclei and of symmetric nuclear matter, which lets us calibrate the model and then extrapolate to subnuclear densities and arbitrary proton fraction. Although our discussion focused on the qualitatively different aspects of the liquid–gas transition in binary systems, the quantitative numerical results are obtained with an equation of state that is as accurate as any currently available. By studying reasonable variations in the nuclear matter symmetry energy and compressibility, we found that these are unlikely to have significant qualitative impact on the signals of the phase transition.

This thermodynamic mean-field model was then used to study the warm nuclear matter that will be produced in energetic heavy-ion collisions. Although the assumption of thermodynamic equilibrium oversimplifies the collision dynamics, we believe it is useful for providing a concrete description of the bulk properties of the warm matter and for examining qualitative features that should have remnants in more microscopic calculations. For example, there are several significant differences between the phase diagram for an asymmetric system and that for symmetric matter. First, the critical temperature $T_c$ is different for different values of the proton fraction $y$, and for $y \neq 0.5$, $T_c$ is not the maximal temperature at which the phase separation can occur. Second, the most extensive spinodal determining the instability boundary in an asymmetric system is determined by a diffusive (chemical) instability, rather than by a mechanical instability signaled by a negative compressibility or by a thermal instability signaled by a negative heat capacity. Thus, in general, *the system is unstable to isovector modes of separation rather than to isoscalar modes.* We also observed that in symmetric matter, the isothermal spinodal determines the region of instability for either isothermal or adiabatic expansion. (The diffusive and isothermal spinodals coalesce

as $y \to 0.5$.)

Finally and most importantly, the dimensionality of the phase-separation region is larger in an asymmetric binary system, which implies that the phase transition is *continuous*, and which allows the thermodynamic properties of the phases to change throughout the transition. Although the changes in the thermodynamic variables are small for realistically observable systems ($0.3 \leq y \leq 0.5$) in the model studied here, the resulting spread in the thermodynamic variables will increase the variations expected from a study of statistical fluctuations alone. Moreover, the increased dimensionality of the phase-separation problem could generate significant changes in the observables calculated in a more microscopic treatment, for example, one based on fluctuations and spinodal decomposition. At the least, one must allow for different proton concentrations in the liquid and vapor when dealing with asymmetric systems, rather than including only fluctuations in the density. Such microscopic calculations remain to be performed and represent important topics for future work on this problem.

## ACKNOWLEDGMENTS

We are pleased to thank our colleagues G. F. Bertsch, J. F. Dawson, R. J. Furnstahl, S. V. Gardner, C. J. Horowitz, and H-B. Tang for useful discussions and comments. This work was supported in part by the Deutscher Akademischer Austauschdienst (DAAD) and by the U.S. Department of Energy under contract No. DE–FG02–87ER40365.

## APPENDIX: THE LOW-DENSITY LIMIT

Here we derive some useful results for computing the low-density limit of the equation of state defined by Eqs. (30) and (47). For densities $\rho \lesssim 0.005\,\mathrm{fm}^{-3}$, the results in Sec. III lead to numerical inaccuracies that can be overcome by making an explicit low-density expansion. In general, one has to distinguish between two limiting cases: $(i)$ $\rho \to 0$ at finite $T$ and $(ii)$ the nonrelativistic limit. We will consider these two cases in turn.

At fixed temperature, zero density requires $\nu_{p,n} \to 0$. We therefore consider

$$\nu_{p,n} \equiv \epsilon \bar{\nu}_{p,n} \quad \text{and} \quad M^* \equiv M_0^* + \epsilon^\alpha \Delta M^* \quad \text{for} \quad \epsilon \to 0 \ ,$$

where $M_0^*$ is the effective mass at zero density. Using the recursion relations in Eqs. (39)–(42), expansions of the densities are readily obtained:

$$\rho = \frac{\epsilon}{\pi^2}(\bar{\nu}_p + \bar{\nu}_n)\left(2H_3(0, M_0^*) + M_0^{*2} H_1(0, M_0^*)\right) + O(\epsilon^2) \ , \tag{A1}$$

$$\rho_3 = \frac{\epsilon}{\pi^2}(\bar{\nu}_p - \bar{\nu}_n)\left(2H_3(0, M_0^*) + M_0^{*2} H_1(0, M_0^*)\right) + O(\epsilon^2) \ , \tag{A2}$$

$$\rho_s = 2\frac{M_0^*}{\pi^2} H_3(0, M_0^*) + O(\epsilon^2) \ , \tag{A3}$$

which gives

$$y = \frac{1}{2}\left(1 + \frac{\bar{\nu}_p - \bar{\nu}_n}{\bar{\nu}_p + \bar{\nu}_n}\right) \quad \text{for} \quad \rho \to 0 \ . \tag{A4}$$

The effective mass $M_0^*$ is determined by the corresponding limit of Eq. (43):

$$\frac{m_\mathrm{s}^2}{g_\mathrm{s}^2}(M - M_0^*) + \frac{\kappa}{2g_\mathrm{s}^3}(M - M_0^*)^2 + \frac{\lambda}{6g_\mathrm{s}^4}(M - M_0^*)^3 = 2\frac{M_0^*}{\pi^2} H_3(0, M_0^*) \ , \tag{A5}$$

which is independent of the proton fraction $y$. By inspecting the lowest-order corrections in $\epsilon$, it follows that $\alpha = 2$, *i.e.*,

$$M^* = M_0^* + O(\rho^2) \ .$$

Moreover, by observing that

$$W \underset{\rho \to 0}{=} O(\rho) \quad \text{and} \quad R \underset{\rho \to 0}{=} O(\rho) \ ,$$

we conclude that all isotherms with $0 < y < 1$ approach a common limit at zero density:

$$\begin{aligned} p \underset{\rho \to 0}{=} \ & \frac{2}{3\pi^2} H_5(0, M_0^*) - \frac{m_\mathrm{s}^2}{2g_\mathrm{s}^2}(M - M_0^*)^2 - \frac{\kappa}{6g_\mathrm{s}^3}(M - M_0^*)^3 \\ & - \frac{\lambda}{24 g_\mathrm{s}^4}(M - M_0^*)^4 + O(\rho^2) \ . \end{aligned} \tag{A6}$$

At moderate temperatures ($T \lesssim M/10$), this vacuum pressure is very small.

We turn now to the nonrelativistic limit, which is encountered in particular in adiabatic processes. This limit can be characterized by

$$\rho \to 0 \quad \text{and} \quad M^*/T \to \infty \ , \quad \text{but} \quad \eta_{p,n} \equiv (\nu_{p,n} - M^*)/T \quad \text{finite.}$$

It is a straightforward task to derive asymptotic expansions for the integrals (37) and (38):

$$G_n(\mu, M) \underset{M/T\to\infty}{=} \frac{(2MT)^{n/2}}{2}\left(F_{n/2-1}(\eta) + (n+2)\frac{T}{4M}F_{n/2}(\eta) + O(T^2/M^2)\right) , \quad \text{(A7)}$$

$$H_n(\mu, M) \underset{M/T\to\infty}{=} T(2MT)^{(n/2-1)}\left(F_{n/2-1}(\eta) + (n-2)\frac{T}{4M}F_{n/2}(\eta) + O(T^2/M^2)\right) , \quad \text{(A8)}$$

with

$$F_\alpha(\eta) \equiv \int_0^\infty dx \frac{x^\alpha}{1+e^{x-\eta}} .$$

To lowest order, this yields the familiar Fermi gas results for the densities:

$$\rho \approx \rho_s \approx \frac{(2MT)^{3/2}}{2\pi^2}\Big[F_{1/2}(\eta_p) + F_{1/2}(\eta_n)\Big] \quad \text{(A9)}$$

and for the entropy

$$\sigma \approx \frac{(2MT)^{3/2}}{6\pi^2}\Big(5\Big[F_{3/2}(\eta_p) + F_{3/2}(\eta_n)\Big] - 3\Big[\eta_p F_{1/2}(\eta_p) + \eta_n F_{1/2}(\eta_n)\Big]\Big) . \quad \text{(A10)}$$

The effective mass in these expressions has been replaced by the nucleon mass $M$, since

$$M^* \approx M - \frac{g_s^2}{m_s^2}\rho , \quad \text{(A11)}$$

which follows from the relation (A9). In an adiabatic process, the constants $\eta_{p,n}$ can be obtained for a given entropy/baryon $s = \sigma/\rho$ and proton fraction $y$ using

$$\frac{\sigma}{\rho} = \frac{5\Big[F_{3/2}(\eta_p) + F_{3/2}(\eta_n)\Big] - 3\Big[\eta_p F_{1/2}(\eta_p) + \eta_n F_{1/2}(\eta_n)\Big]}{3\Big[F_{1/2}(\eta_p) + F_{1/2}(\eta_n)\Big]} , \quad \text{(A12)}$$

$$2y - 1 = \frac{F_{1/2}(\eta_p) - F_{1/2}(\eta_n)}{F_{1/2}(\eta_p) + F_{1/2}(\eta_n)} . \quad \text{(A13)}$$

By including the lowest-order contributions in the density, the pressure then takes the form

$$p(\rho, s, y) \approx \frac{(2\pi^2)^{2/3}}{3M}\rho^{5/3}\frac{F_{3/2}(\eta_p) + F_{3/2}(\eta_n)}{(F_{1/2}(\eta_p) + F_{1/2}(\eta_n))^{5/3}} + \frac{g_v^2}{2m_v^2}\rho^2 - \frac{g_s^2}{2m_s^2}\rho^2 + \frac{g_\rho^2}{8m_\rho^2}(2y-1)^2\rho^2 . \quad \text{(A14)}$$

As expected, at very low densities the pressure is dominated by the ideal-gas term with adiabatic index 5/3. The nonlinear couplings give rise to higher-order contributions in the density and can be neglected.